\documentclass[article]{aa}
\usepackage{graphicx} 
\usepackage{lscape} 
\usepackage{multirow}
\usepackage[caption=false]{subfig} 
\usepackage[switch]{lineno}
\usepackage{placeins}
\usepackage{txfonts}
\begin{document} 

   \title{Period-Luminosity Relations for Galactic Type II Cepheids in the Sloan 
   bands\thanks{Based on data from the Las Cumbres Observatory. 
   The light curves are only available in electronic form at the Araucaria Project webpage: 
   \url{https://araucaria.camk.edu.pl/} and the CDS via anonymous ftp to \url{cdsarc.u-strasbg.fr} 
   (130.79.128.5) or via \url{http://cdsweb.u-strasbg.fr/cgi-bin/qcat?J/A+A/}.}}


   \titlerunning{PL relations for T2Ceps stars in the Sloan bands}
   \authorrunning{W. Narloch et al.}

   \author{W. Narloch\inst{1}, 
          G. Hajdu\inst{1}, 
          G. Pietrzyński\inst{1}, 
          P. Wielgórski\inst{1},
          R. Smolec\inst{1},
          W. Gieren\inst{2}, 
          B. Zgirski\inst{2}, 
          M. Górski\inst{1}, \\
          P. Karczmarek\inst{1} 
          \and
          D. Graczyk\inst{3}
          }

   \institute{Nicolaus Copernicus Astronomical Center, Polish Academy of Sciences, 
             Bartycka 18, 00-716 Warszawa, Poland\\
              \email{wnarloch@camk.edu.pl}
         \and
             Universidad de Concepci\'on, Departamento de Astronom{\'i}a, 
             Casilla 160-C, Concepci\'on, Chile
         \and
             Nicolaus Copernicus Astronomical Center, Polish Academy of Sciences, 
             Rabia\'nska 8, 87-100 Toru\'n, Poland
             }

   \date{Accepted: 10th March 2025}

 
  \abstract
   {Type~II Cepheids (T2Ceps), alongside RR~Lyrae stars, serve as important distance 
   indicators for old population~II stars due to their period-luminosity (PL) relations. 
   However, studies of these relations in the Sloan photometric system are rather limited 
   in the literature.}
   {Our goal is to calibrate PL relations (and their counterparts in Wesenheit magnitudes) 
   in the Sloan--Pan-STARRS $g_{P1}r_{P1}i_{P1}$ bands for Galactic T2Ceps located in the 
   vicinity of the Sun.}
   {We collected data for $16$ T2Ceps of the BL~Her type and $17$ of the W~Vir type using 
   $40$~cm telescopes of the Las Cumbres Observatory Global Telescope Network. Geometric 
   parallaxes were adopted from Gaia Data Release $3$.}
   {We have calibrated PL and period-Wesenheit relations for Milky Way BL~Her and W~Vir 
   stars in the solar neighborhood, as well as for a~combined sample of both types.}
   {The relationships derived here will allow to determine the distances to T2Ceps 
   that will be discovered by the Legacy Survey of Space and Time survey and, in turn, to 
   probe the extended halo of the Milky Way, as well as the halos of nearby galaxies. To the 
   best of our knowledge, the relations derived in this study are the first for Milky Way 
   T2Ceps in the Sloan bands.}
            

   \maketitle
%

\section{Introduction} \label{sec:intro}

   Cepheid variables are pulsating stars known for their usefulness as distance indicators, 
   thanks to their period-luminosity (PL) relations, first discovered by Henrietta Swan Leavitt 
   \citep[][also known as the Leavitt law]{LP1912}. Further studies of these variables led 
   to a~discovery of hydrogen emission lines in spectra of the Cepheid W~Virginis, which somehow 
   distinguished it among other Cepheids \citep{Joy1937}. \citet{Joy1940} reported changes in 
   the spectrum of the Cepheid V154 in the globular cluster M3 resembling those in W~Virginis 
   star. 
   In 1944 Walter Baade \citep{Baade1944} resolved stars in the center of the M31 for the first 
   time. These observations led him to distinguish between two stellar populations, known today 
   as Population~I and II. Later, he also discovered \citep{Baade1956} the existence of two types 
   of Cepheid variables, each having different PL relation. Population~II Cepheids were found to 
   be approximately $1.5$~mag fainter in blue or visual compared to Population~I Cepheids of the 
   same period. The separation of these groups of pulsators resulted in a~doubling of the 
   estimated distances to nearby galaxies, thereby revising the cosmic distance scale. 
   Today, the terms Type I (or classical) Cepheids and Type II Cepheids (hereafter T2Ceps) 
   are used to describe these two types of variables. They occupy different regions of the 
   classical instability strip, reflecting their different evolutionary and physical properties. 
   Type~II Cepheids are old, evolved stars less massive than the Sun, and because of that they 
   serve as distance indicators to old stellar populations, alongside other classical pulsators 
   RR~Lyrae stars. In contrast, Type I Cepheids, which are young radially pulsating 
   massive evolved giants, serve as distance indicators to young stellar populations. 
   Their locations within the Milky Way (MW) are also different. Type~I Cepheids are mostly 
   found in the Galatic disk, and are strongly associated with the spiral arms 
   \citep{Skowron2019}. Meanwhile, Type~II Cepheids are commonly found in old stellar 
   populations like the Galactic bulge \citep{Dekany2019} and halo \citep{Matsunaga2006}.
   
   T2Ceps are divided into subgroups depending on the distribution of their pulsation period, 
   although this division is not strict and depends on the environment 
   \citep[e.g.,][]{Gingold1985,Soszynski2011}. 
   First division of T2Ceps might be date back to 1949 \citep{Joy1949}. Joy distinguished two 
   groups of short period (with periods between $1-3$~days) and long period T2Ceps ($13-19$~days), 
   where the second ones resembled W~Virginis Cepheid. The justification for such a~division was 
   the existence of a~gap in the periods between $5-10$~days, very poorly populated by stars. 
   This division has persisted to the present day, where the first group is now known as BL~Her 
   type stars, and the second group as W~Wir type stars 
   \citep[e.g.,][]{Soszynski2008,Soszynski2011}. 
   The second class smoothly transitions into a~third group of stars with even longer periods 
   (above about $20$~days), known as RV~Tau type stars, whose characteristics include different 
   depths of consecutive light minima (period doubling). 
   It is believed that these three subclasses are similar stars on different stages of stellar 
   evolution. 
   \citet{Soszynski2008} distinguished another class of T2Ceps on average brighter and bluer than 
   regular W~Vir stars, also having different shapes of the light curves, and called them peculiar 
   W~Vir stars (hereafter pW~Vir). The explanation for their peculiarity is binarity, as indicated 
   by the ellipsoidal or eclipsing variability often found in their light curves. 

   PL relations of T2Ceps are extensively investigated in many bands, both theoretically 
   \citep[e.g.,][]{DiCriscienzo2007,Das2021,Das2024} and empirically. Of particular interest 
   to the community are the PL relations in the near-infrared (NIR) bands, because of the low 
   dependence on the reddening. The NIR PL relations are derived in different environments 
   containing old stellar populations in the MW, such as Galactic bulge 
   \citep[e.g.,][]{Groenewegen2008,Bhardwaj2017b,Braga2018a}, Galactic field 
   \citep[e.g.,][]{Wielgorski2022},  
   Galactic globular clusters \citep[e.g.,][]{Matsunaga2006,Bhardwaj2022}, 
   as well as nearby galaxies 
   \citep[e.g.,][]{Matsunaga2009,Matsunaga2011,Ripepi2015,Bhardwaj2017a,Bhardwaj2022,Sicignano2024}. 
   But there exist also a~number of works in optical range 
   \citep[see, e.g.,][]{Alcock1998,GJ2017,Iwanek2018,Ripepi2023}. 
   In the wide-band Sloan photometric filters \citep[$ugriz$,][]{Fukugita1996}, however, there 
   are very few studies, namely \citet{Ngeow2022T2Cep} in Galactic globular clusters and 
   \citet{Kodric2018} in M31, both in the Pan-STARRS version of the Sloan photometric system 
   \citep{Tonry2012}. 
   In particular, there are no PL relations determinations based on T2Cep variables from the 
   vicinity of the Sun in those bands, which we endeavour to change in this study.  
   
   We aim to establish the PL relations and their counterparts in Wesenheit magnitudes 
   \citep{Madore1982} for Galactic T2Ceps from the neighborhood of the Sun, using the Sloan 
   $gri$ bands calibrated specifically to the Pan-STARRS implementation of this photometric 
   system, which will be used, among others, in the upcoming $10$~yrs Vera C. Rubin Observatory 
   Legacy Survey of Space and Time \citep[Rubin-LSST;][]{Ivezic2019}. 
   Therefore, providing accurate calibrations of the PL relations within this photometric system 
   is of great interest. 
   This study complements our earlier investigations on similar relations for other types of 
   pulsating stars, namely classical Cepheids \citep{Narloch2023} and RR~Lyrae stars 
   \citep{Narloch2024}. 
   To the best of our knowledge, existing PL relations in the Sloan bands for T2Ceps in the MW 
   have so far been derived only for stars in globular clusters \citep{Ngeow2022T2Cep}, thus, 
   our relations will be the first of their kind. We believe that this work will serve as 
   a~valuable contribution to the field and a~practical tool for measuring distances in the 
   universe in the era of large-scale sky surveys like Rubin-LSST. 
   
   This paper is organized as follows. Section~\ref{sec:data} provides information about the 
   sample of Galactic T2Ceps observed for the project, the collected scientific data, their 
   reduction, and the methods used to determine the absolute magnitudes of our stars. In 
   Sect.~\ref{sec:relations}, we describe the derivation of the PL and period-Wesenheit (PW) 
   relations and explain the reasons for rejecting specific stars from the fitting procedure. 
   Section~\ref{sec:discussion} provides a~short discussion of the results, while 
   Sect.~\ref{sec:summary} concludes the paper.  


\section{Data} \label{sec:data}


\subsection{Sample of stars and classification} \label{ssec:sample}


   \begin{figure*}
   \centering
   \includegraphics{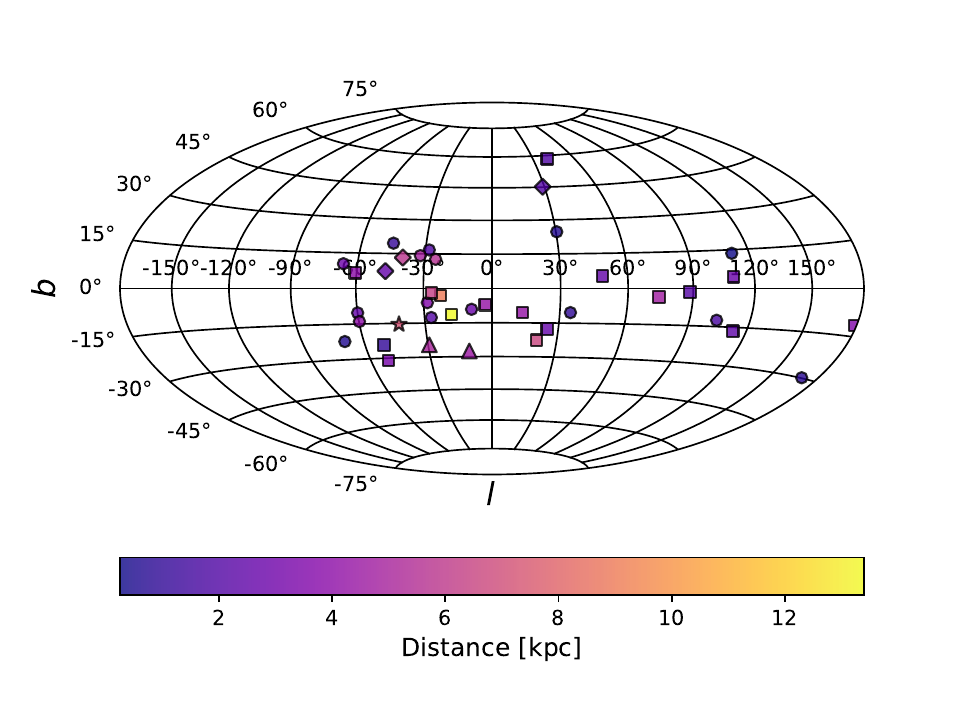}
   \caption{Location of the Type~II Cepheids (BL~Her type: circle, W~Vir type: squares) used 
            for establishing the PL relations in this paper given in Galactic coordinates, 
            as well as other types of variable originally observed for the project and 
            reclassified later (pW~Vir type: diamonds, RV~Tau type: triangles, Type~I 
            first-overtone Cepheid: star). 
            \label{fig:galmap}}
    \end{figure*}
   The original sample of T2Ceps was selected from lists of variable stars used for other 
   projects of the Araucaria group \citep{Gieren2005}. We observed a~total of $17$ BL~Her and 
   $22$ W~Vir type stars. 
   However, in a~later Fourier analysis of the light curves (presented in Appendix~\ref{app:fp}), 
   V572~Aql turned out to be a~Type~I Cepheid pulsating in the first overtone mode. Additionally, 
   three stars (AL~Vir, AP~Her, and BH~Oph) were classified as pW~Vir stars based on the shape 
   of their light curves, while TW~Cap and V1711~Sgr were identified as RV~Tau type stars. 
   Eventually, we were left with 16 BL~Her type stars and 17 W~Vir type stars for calibrating 
   the PL relations.  
   
   The range of Gaia $G$-band magnitudes for BL~Her type stars is between $7.11$ and $13.71$~mag 
   (with a~mean of about $11.20$~mag), and for W~Vir type stars between $9.04$ and $13.50$~mag 
   (with a~mean of about $11.64$~mag). 
   The distances derived as the inverse of Gaia Data Release~$3$ (DR3) parallaxes 
   \citep{GaiaCol2023}, corrected for zero point (ZP) offsets as proposed by 
   \citet{LindegrenBastian2021}, for BL~Her stars range from about $0.3$ to $5.6$~kpc (with 
   a~median of about $2.0$~kpc), and for W~Vir, the range is from about $1.1$ to $13.4$~kpc 
   (with a~median of about $3.6$~kpc). Figure~\ref{fig:galmap} shows the sky distribution of the 
   analyzed stars.
   
   Pulsation periods were adopted from the International Variable Star 
   Index\footnote{\url{https://www.aavso.org/vsx/}} (AAVSO), and they range between $1.09$ 
   and $3.95$~days for BL~Her, and $6.17$ to $19.95$~days for W~Vir type stars.  
   Table~\ref{tab:t2cep} lists the parameters of all the stars in the original sample. 
   For HQ~Cen and RS~Pav, with alternating deep and shallow minima, we adopted half of the 
   period given by AAVSO. 
   We marked the periods of these stars in Table~\ref{tab:t2cep} with double asterisk.
   

\subsection{Data and reduction} \label{ssec:data_reduction}

   
   Data for the project were collected from August $2021$ to July $2022$ using $14$ robotic 
   $40$~cm telescopes from the Las Cumbres Observatory (LCO) Global Telescope 
   Network\footnote{\url{https://lco.global/}}. Observations were conducted under the 
   programs CLN2021B-008 and CLN2022A-008. 
   Images were taken with Sloan $g'r'i'$ filters using 3K$\times$2K SBIG STL-6303 cameras, 
   which provided a field of view of $29.2\times19.5$ arcmin$^{2}$ and a~pixel size of 
   $0.571$~arcsec per pixel, with no binning. The air mass of the observations ranged from 
   $1.0$ to $1.6$, and the average seeing was approximately $2.40$, $2.32$, and $2.36$~arcsec 
   in the Sloan $g',\ r'$ and $i'$ bands, respectively.   
   
   We retrieved pre-reduced and processed images from the 
   LCO Archive\footnote{\url{https://archive.lco.global/}}, which had been processed with the 
   LCO BANZAI\footnote{\url{https://lco.global/documentation/data/BANZAIpipeline/}} pipeline. 
   Aperture photometry and data calibration followed the procedures outlined in 
   \citet{Narloch2023}, which we recommend for detailed methodology. 
   The instrumental magnitudes obtained with the standard DAOPHOT package \citep{Stetson1987} 
   were cross-matched with the ATLAS All-Sky Stellar Reference Catalog version~$2$ 
   \citep[ATLAS-REFCAT2;]{Tonry2018}, which is on the Pan-STARRS version of the Sloan photometric 
   system \citep{Tonry2012}, referred to as $g_{P1}r_{P1}i_{P1}$ later in the text. 
   The mean DAOPHOT photometric uncertainties were about $0.04$~mag and $0.03$~mag for $g_{P1}$ 
   and $r_{P1}i_{P1}$ filters, respectively, for stars with magnitudes in the range of 
   $9.0 < g_{P1} < 15.0$~mag. 
   This range encompasses the magnitudes of our sample stars and most of the reference stars. 
   For the photometric calibration, we used the equations presented in \citet{Narloch2023}, which 
   were selected in such a~way as to minimize the influence of the nonlinearity of the LCO SBIG 
   cameras. We note that this effect was less severe for T2Ceps when compared to classical 
   Cepheids, as most of the reference stars had brightness levels similar to that of the target 
   star. 
   
   The intensity-averaged mean apparent magnitudes were obtained by fitting Fourier series 
   to the light curves presented in Appendix~\ref{app:lc} 
   (Figs.~\ref{fig:fig1} -- \ref{fig:fig3}), and are given in Cols.~$8-10$ of 
   Table~\ref{tab:t2cep}. 
   The orders of the Fourier series ranged from $2$ to $9$ and were chosen to best fit the shape 
   of the light curves. For those of W~Vir and RV~Tau type stars, which exhibit varying depths 
   of minima (marked in Table~\ref{tab:t2cep} with single and double asterisks), the mean 
   magnitudes were calculated by fitting a~Fourier series with double their formal periods, 
   while on the PL relations they were marked with single periods, which are intervals between 
   successive minima. 


\subsection{Reddening} \label{ssec:redd}


   We corrected the mean magnitudes of our stars using the E(B-V) color excess values from the 
   reddening map of \citet[][hereafter SF map]{SF2011}, integrated up to the distance 
   of the target stars with the assumption of the three-dimensional MW model of 
   \citet{DrimmelSpergel2001} \citep[see][for details of adopted parameters]{Suchomska2015}. 
   The values used are given in Col.~$7$ of Table~\ref{tab:t2cep}. 
   The extinction vectors ($R_{\lambda}$) for the Pan-STARRS $g_{P1}r_{P1}i_{P1}$ bands were 
   adopted from \citet[][see their Table~1]{Green2019}, with values of $R_{g} = 3.518$, 
   $R_{r} = 2.617$ and $R_{i} = 1.971$. These values were used to calculate three Wesenheit 
   indices, which are inherently reddening-free for a~specific reddening law \citep{Madore1982}. 
   We defined the Wesenheit indices as follows: $W^{ri}_r = r - 4.051 (r-i) - \mu$, 
   $W^{gr}_r = r - 2.905 (g-r) - \mu$, and $W^{gi}_g = g - 2.274 (g-i) - \mu$, where $\mu$ 
   represents the distance modulus (DM).


\subsection{Distances} \label{ssec:dist}


   To calculate the absolute magnitudes ($M_{\lambda}$) of stars, necessary for calibrating 
   the PL relations, from the dereddened mean apparent magnitudes ($m_{\lambda}$), one needs 
   to know the distances of the target stars. 
   Analogously to \citet{Narloch2024}, we applied four different approaches to accomplish this: 
   the parallax method, the Astrometry-Based Luminosity (ABL) method 
   \citep{FC1997,AL1999}, and finally geometric and photogeometric distances provided by 
   \citet{BJ2021}. The applied equations used to calculate the absolute magnitudes in each 
   method can be found in \citet{Narloch2024}.

   The range of Gaia~DR3 parallaxes of our sample stars, corrected for the ZP offset calculated 
   as proposed by \citet{LindegrenBastian2021}\footnote{The parallax ZP offset corrections for 
   each star were derived using the dedicated Python code: 
   \url{https://gitlab.com/icc-ub/public/gaiadr3_zeropoint}.} 
   by taking into account the ecliptic latitude, 
   magnitude, and color of a~star, was between $0.18$ and $3.97$~mas for BL~Her stars (with 
   a~median of $0.50$~mas), and about $0.07$ to $0.94$~mas for W~Vir stars (with a~median of 
   about $0.28$~mas). They are provided in Col.~$4$ of Table~\ref{tab:t2cep}. The range 
   of applied ZP offset corrections was from $-15$ to $-44\,\mu$as for BL~Her stars (with a~mean 
   of about $-30\,\mu$as), and from $-4$ to $44\,\mu$as for W~Vir stars (with a~mean of about 
   $-25\,\mu$as). 
   \citet{LindegrenBastian2021} recommended including an uncertainty of a~few $\mu$as in the ZP; 
   therefore, we adopted $5\,\mu$as as the systematic error. 
   
   The quality of the parallaxes was evaluated using two parameters: RUWE and GOF given in the 
   Gaia~DR3 catalog. The RUWE parameter is sensitive to the photocentric motion of unresolved 
   objects, such as astrometric binaries.  
   Meanwhile, GOF can be used as an indicator of the level of asymmetry of a~source 
   \citep[e.g.,][]{Riess2021}. Based on the criteria from \citet{Breuval2021} and 
   \citet{Wielgorski2022}, we excluded stars with $\rm{RUWE}>1.4$ or $\rm{GOF}>12.5$ from our 
   T2Ceps sample. 
   One star of BL~Her type (V465~Oph) was rejected based on the RUWE criterion, along with three 
   W~Vir type stars (AL~Lyr, ST~Pup and TX~Del). Additionally, one RV~Tau type star (V1711~Sgr) 
   did not meet this parallax quality standard. Among those stars, only W~Vir type stars had 
   $\rm{GOF}>12.5$, making RUWE the stricter criterion in our case.  
   
   The mean parallax uncertainty for T2Ceps from our sample (given in Col.~$4$ of 
   Table~\ref{tab:t2cep}) is approximately $19\,\mu$as. 
   During the calculations, we found that one of the W~Vir type stars falls within the problematic 
   magnitude range of $G = 11.0 \pm 0.2$~mag, where the transition between Gaia window classes 
   may impact the parallax ZP \citep[see Fig.~$1$ in][]{LindegrenBastian2021}. 
   Another two BL~Her type stars fall within the range of $G = 12.0 \pm 0.2$~mag. We quadratically 
   added $10\,\mu$as to the parallax uncertainties of those stars, thereby following 
   \citet{Breuval2021}, to account for potential additional errors caused by the transition 
   windows. Furthermore, following the recommendation of \citet{Riess2021}, we eventually 
   increased all Gaia~DR3 parallax uncertainties by $10\%$ to account for potential additional 
   uncertainties.


\section{Derived relations} \label{sec:relations}



\subsection{PL and PW relations in the Sloan bands} \label{ssec:PLPWrel} 


   We used absolute Sloan-Pan-STARRS $g_{P1}r_{P1}i_{P1}$ magnitudes, derived as described in 
   previous sections, for the calibration of the PL relation, defined as: 

   \begin{equation} \label{eq:plr}
     M_{\lambda} = a_{\lambda} (\log P - \log P_0) + b_{\lambda},
   \end{equation}

   \noindent as well as analogous relations for Wesenheit magnitudes (derived in 
   Sect.~\ref{ssec:redd}), defined as follows: 
   
   \begin{equation} 
     W = a (\log P - \log P_0) + b \label{eq:pwr}, 
   \end{equation} 

   \noindent where $a_{\lambda}$ and $b_{\lambda}$ (and $a$ and $b$, analogously) are the 
   sought-after slope and intercept, respectively. 
   In order to minimize correlation between the two parameters, we subtracted the logarithm of 
   a~pivot period ($\log P_0$). Following \citet{Wielgorski2022}, we adopted $\log P_0 = 0.3$ 
   for BL~Her type stars, while for W~Vir stars we chose $\log P_0 = 1.2$. 
   \citet{Soszynski2008} indicate that the PL relations for T2Ceps are not linear across the 
   entire range of periods, so they rather should be fitted separately for BL~Her, W~Vir, and 
   RV~Tau stars. 
   However, the nonlinear behavior is particularly characteristic for RV~Tau type stars, while 
   the other two classes of T2Ceps exhibit linear behavior. 
   On the other hand, \citet{Sicignano2024} emphasize that the three subclasses of T2Ceps do not 
   follow unique PL relation particularly in optical bands, while they start to do that in the 
   bands redder than $I$-band. This could be related with the location of specific T2Cep types 
   in the instability strip in the optical bands. 
   Nevertheless, we also derived coefficients for a~combined sample of BL~Her+W~Vir stars 
   \citep[as often is done in the literature, e.g.,][]{Ngeow2022T2Cep}, for 
   which we adopted $\log P_0 = 0.7$, similarly to \citet{Wielgorski2022}. 
   {To test if BL~Her and W~Vir stars can be treated as one sample, we performed 
   the F-Snedecor statistical test, which checks the equality of two variances. 
   We compared the variances for a~combined sample where the two populations of stars were fitted 
   with separate and global PL/PW relations. 
   We found our F-values to be smaller than the critical values for degrees of 
   freedom of $23$ (PL relations) and $25$ (PW relations) at a~significance level of 
   $\alpha = 0.05$. The interpretation of this result is that the obtained values of standard 
   deviations do not differ from each other in a~statistically significant manner (i.e., the 
   compared methods do not differ in terms of precision). Therefore, there is no basis for 
   rejecting the hypothesis that BL~Her and W~Vir stars can be fitted together. 
   However, taking into consideration the lack of consensus on this matter existing in the 
   literature, we decided to calibrate separate PL/PW relations for BL~Her and W~Vir stars, 
   as well as a~single relation for the combined sample.
   
   For the absolute magnitudes calculated with the ABL method, 
   we fitted the following relation: 
   
   \begin{equation} \label{eq:abl_plr} 
     ABL_{\lambda} = 10^{0.2[a_{\lambda}(\log P - \log P_0) + b_{\lambda}]}. 
   \end{equation}
   
   \noindent and analogously for Wesenheit magnitudes in the form: 
   
   \begin{equation} \label{eq:abl_pwr} 
     ABL_{W} = 10^{0.2[a(\log P - \log P_0) + b]}. 
   \end{equation} 
   
   Similarly to \citet{Narloch2023}, we used the {\tt curve\_fit} function from the {\tt scipy} 
   Python library to perform the fitting, followed by a~$3\sigma$-clipping procedure. For the 
   final statistical uncertainties of the resulting coefficients, we adopted the errors returned 
   by the fitting function. 
   The derived coefficients for all four methods are presented in Table~\ref{tab:plr}. 
   Figures~\ref{fig:plr_blher+wvir} and \ref{fig:pwr_blher+wvir} present resulting 
   PL and PW relations, respectively, based on the parallax and ABL methods only. The error 
   bars for each star are derived from the error propagation taking into account statistical 
   mean magnitudes and parallaxes errors. 
      
   In the final fitting of PL relations we used $14$ stars of BL~Her stars and $10$ stars of 
   W~Vir type ($24$ for the combined sample). In case of the PW relations these were $15$ and 
   $11$ stars, respectively, which gave $26$ stars in the combined sample. 
   Some stars have been manually excluded from the fit. These will be discussed in the next 
   Sect.~\ref{ssec:rej}. 
   The rms of the PL relations were between $0.20 - 0.24$, $0.32 - 0.39$, and $0.26 - 0.31$ for 
   BL~Her, W~Vir and combined sample, respectively. For PW relations these numbers are 
   $0.19 - 0.23$, $0.28 - 0.33$, and $0.24 - 0.27$, respectively.


\subsection{Stars excluded from the fit} \label{ssec:rej} 


   One star among BL~Her type variables, AU~Peg, deviates towards fainter brightnesses on the 
   PL relation. This star was excluded by \citet{Wielgorski2022} from their NIR PL relations, 
   where they also noted its unusually red color ($J-K_{\mathrm{s}} \approx 0.7$~mag). Our data 
   confirm this observation (see Fig.~\ref{fig:cmd}), showing that this star stands out 
   in color compared to other variables of this type. Consistent with \citet{Wielgorski2022}, we 
   also observe that the deviation from the fitted PL relation is more pronounced in bluer bands 
   and disappears in Wesenheit indices, suggesting problems with reddening for this star. Although 
   the reddening value listed in Table~\ref{tab:t2cep} is relatively low, there are grounds to 
   believe it may be underestimated. 
   AU~Peg has been identified as a~single-lined spectroscopic binary with a~likely more massive 
   companion \citep{Harris1979}, a~short orbital period of $P_{orb} \approx 53$~days 
   \citep{Harris1984}, and shows rapid period changes that may be due to the companion star 
   \citep[e.g.,][]{Vinko1993}. 
   \citet{McAlaryWelch1986} postulated the presence of relatively thick circumstellar matter 
   around the system. This circumstellar cloud might be responsibile for the unusual color of 
   AU~Peg. 
   As an additional test, we adopted reddening values from other sources to check if 
   they improve the position of AU~Peg in the PL relation. We used the STructuring by Inversion 
   the Local Interstellar Medium (STILISM) map\footnote{\url{https://stilism.obspm.fr/}} 
   \citep{Lallement2014,Capitanio2017}, the {\tt Bayestar2019} 3D reddening map 
   \citep{Green2019}, accessed through the {\tt dustmap} 
   code\footnote{\url{https://dustmaps.readthedocs.io/en/latest/}}, and the SFD reddening 
   map from \citet{SFD1998}. However, for AU~Peg, the reddening values are very similar 
   (SF: $0.046$~mag, SFD: $0.054$~mag, STILISM: $0.069$~mag, {\tt Bayestar2019}: $0.085$~mag), 
   and the new values do not change its position in the PL relations much, so the problem 
   persists.  
   We therefore decided to exclude this star from PL relation fit but retained it in the PW 
   relation fit.  
   
   Some stars of W~Vir type deviated from the obtained relations, even after $3\sigma$-clipping 
   procedure, so we decided to reject them manually from the fit. The first such example is 
   DD~Vel, which showed significant deviation in the PL relation, but not in the PW relation. 
   Additionally, it deviates more towards the blue and less in the red filters, which is an 
   indication of an incorrect value of the adopted reddening, rather than the misclassification 
   of this star as a~Type II Cepheid, which is quite certain \citep[see][]{Lemasle2015}. 
   Figure~\ref{fig:cmd}, where DD~Vel is shifted significantly towards the bluer colors, also 
   strongly suggests that the star is over-derredened. 
   Indeed, the E(B-V) value for this star in Table~\ref{tab:t2cep} is suspisiously large and 
   most likely incorrect. It also lies above the NIR PL relationships of 
   \citet[][see their Fig.~18]{Wielgorski2022}, where they used the \citet{SF2011} maps 
   as the source of the reddening values as well. 
   Again, we tested the reddening values from other reddening maps, as we 
   did for AU~Peg in the previous paragraph. The reddening values from the SF and SFD maps are 
   similarly large (SF: $1.334$~mag, SFD: $1.551$~mag), while the STILISM map gives a~value of 
   $0.201$~mag which, for a~change, places DD~Vel below the PL relation, strongly suggesting 
   that this is also an incorrect value. On the other hand, the {\tt Bayestar2019} map does not 
   provide a~reddening value for this target at all.
   Therefore, we rejected DD~Vel from the fit of the PL relations but kept it in the PW relations. 

   \citet{Wielgorski2022} noticed that the~star QX~Aqr lies below their NIR PL relations. The 
   same is true in our optical data. The reason for this behavior is unknown, so we decided to 
   manually reject this star from the fit of the PL and PW relations. 
   In their work, they also report that the prototype of the class, W~Vir itself lies below the 
   NIR PL relation as well, but it does not appear to deviate from the relations in the Sloan 
   bands. Hence, we kept this star in the fitting process.  
   
   The star V410~Sgr deviated from the PL/PW relations towards brighter magnitudes. Upon 
   investigating its images, we noticed that it has a~close neighboring star that falls within 
   the aperture used for the photometry. To address this, we corrected the photometry for this 
   star by using a~smaller aperture \citep[$8.7$~pixels, corresponding to about 
   $5$~arcsec, instead of $14$~pixels as described in][]{Narloch2023}. 
   The new magnitude of V410~Sgr, however, was only about $0.2$~mag fainter, which still does 
   not bring the star close to the PL/PW relations. In addition, V410~Sgr turns out to be the 
   most distant star in our sample (see Table~\ref{tab:t2cep}), and although the parallax error 
   is rather small, this value itself might be incorrect. An overestimated distance of the object 
   translates into an absolute magnitude that is too bright, which is also suggested by the 
   color-magnitude diagram from Fig.~\ref{fig:cmd} (right panel). 
   The Gaia~DR4 release may potentially resolve this issue in the future. 
   For the time being, we have decided to manually exclude this star from the fitting process.  
   
   Similarly, CO~Sct deviated significantly from the relationships, particularly the PW relations. 
   It also seems to be too bright on the color-magnitude diagram (Fig.~\ref{fig:cmd}, 
   right panel). 
   It is the second most distant star in our sample (see Table~\ref{tab:t2cep}), having 
   simultaneously the largest parallax error among our stars. Therefore, we decided to manually 
   remove it from the fit as well.
   
   Three stars classified as pW~Vir indeed are clustered above determined relations, which is 
   characteristic of these stars and seems to confirm their classification also based on PL/PW 
   relations in addition to Fourier parameters. One RV~Tau type star and the first-overtone Type~I 
   Cepheid are also in locations on the period-luminosity plots typical of their variability 
   types.
   

\section{Discussion} \label{sec:discussion} 

   

\subsection{The influence of the parallax ZP on the PL/PW relations} \label{ssec:ZPinfluence} 


   The Gaia parallaxes presented in Table~\ref{tab:plr} and used to determine the PL/PW 
   relationships through both classical and ABL methods (see Sect.~\ref{ssec:dist}) have 
   been corrected for the ZP offset based on \citet{LindegrenBastian2021}. 
   Geometric and photogeometric distances from \citet{BJ2021} already incorporate these 
   corrections. In line with our previous works  \citep{Narloch2023,Narloch2024}, we 
   examined the impact of these applied corrections by comparing the differences in ZPs between 
   T2Cep PL/PW relations derived for the two distinct cases: correcting Gaia parallaxes using 
   the ZP offset obtained from prescription of \citet{Groenewegen2021} and leaving them 
   uncorrected.   
   
   The mean parallax corrections calculated using the recipe from \citet{Groenewegen2021} are \
   approximately $-28\,\mu$as and $-25\,\mu$as for BL~Her and W~Vir stars, respectively, closely 
   match the values from \citet[][see Sect.~\ref{ssec:dist}]{LindegrenBastian2021}. That implies 
   minor differences in the recalculated PL/PW relation coefficients, with slopes and intercepts 
   well within $1\sigma$ uncertainty. 
   In the case where no parallax corrections are applied, the slopes remain within $1\sigma$ 
   uncertainty, however, the intercepts differ more significantly, resulting in smaller ZPs. 
   These shifts amount to approximately $0.16$~mag for BL~Her stars and $0.29$~mag for W~Vir stars 
   (about $0.20$~mag for the combined sample). The ZP discrepancy between the two types of 
   variables may stem from the fact that W~Vir stars in our sample are generally at greater 
   distances than BL~Her stars. The intercept shifts observed in T2Ceps are considerably larger 
   than those seen in classical Cepheids \citep{Narloch2023} or RR~Lyr stars \citep{Narloch2024}. 
   However, it is noteworthy that, particularly in the latter case, all stars were located 
   at much closer distances (within $3$~kpc). 


\subsection{Comparison of the PL/PW relations with the literature} \label{ssec:comp_with_lit} 


\subsubsection{Comparison with the PL/PW relations in the Galacitc globular clusters} 
\label{ssec:comp_with_Ngeow}

   \citet{Ngeow2022T2Cep} published PL/PW relations for T2Ceps from $18$ globular clusters 
   in the $gri$~bands, based on the Sloan--Pan-STARRS version of the Sloan photometric system. 
   They as well used the reddening vectors from \citet{Green2019} to derreden the brightnesses 
   and calculate Wesenheit magnitudes. Therefore, we can directly compare their relationships 
   with our findings derived using parallax method, as shown in 
   Fig.~\ref{fig:comp_Ngeow2022}. 
   The slopes of our relations are slightly flatter than those presented by 
   \citet{Ngeow2022T2Cep}. 
   Nevertheless, they agree well within about $1\sigma$ uncertainty.  
   However, there is a~significant disagreement in the ZPs of the relationships on the level 
   of approximately $0.2 - 0.4$~mag. This discrepancy appears to be more pronounced in the PL 
   relations than in the PW relations, suggesting that reddening might be a major contributing 
   factor (but certainly not the only one) of an existing difference. To test this hypothesis, 
   we adopted reddening values for the T2Ceps in our sample from a~different sources to 
   investigate    their influence on the ZPs. The advantage of using \citet{SF2011} reddening 
   map is that it provides values for each of our targets, unlike other maps, which may have 
   limited distance ranges or incomplete sky coverage. 
   We used the STILISM map to obtain reddening values of our T2Ceps.
   For stars located at distances not covered by the map, we adopted the maximum reddening value 
   in their direction. 
   For comparison purposes, we also used the {\tt Bayestar2019} 3D reddening map.
   For stars without available reddening values, we assumed zero reddening. The PL relations 
   derived using these new $E(B-V)$ values are shown in Fig.~\ref{fig:comp_Ngeow2022} 
   with green 
   and blue dashed lines. These relations are consistent with each other and with our original 
   results, although yield slightly flatter slopes. However, they exhibit even worse agreement 
   with the relations from \citet{Ngeow2022T2Cep}, ruling out the adopted reddening as the primary 
   source of the disagreement. 
   It is also important to note at this point that the relationships from \citet{Ngeow2022T2Cep} 
   were derived using BL~Her, W~Vir, and RV~Tau stars, thus covering a wider range of periods.  
   In contrast, our calibrated relations for combined sample omit the RV Tau stars, which may 
   contribute to the observed differences. 
   Another issue that may have a~significant impact on this comparison is the range of 
   metallicity of the T2Ceps from our sample compared to the sample from \citet{Ngeow2022T2Cep}.
   
\subsubsection{Application to M31 T2Ceps} 
\label{ssec:comp_with_Kodric}
   
   \citet{Kodric2018} provided average magnitudes in the Sloan--Pan-STARRS bands for $278$ T2Ceps 
   in the galaxy M31, which we used to test our calibrated relations. However, it is important 
   to note that this comparison is problematic for a~few reasons. First, the majority of T2Ceps 
   measured by \citet{Kodric2018} are of the RV~Tau type, while the remaining $68$ stars could 
   be classified as W~Vir stars \citep[assuming a~$20$~day period as the limit between W~Vir 
   and RV~Tau stars; see, e.g.,][]{Soszynski2011}. RV~Tau type stars exhibit nonlinear behavior 
   in optical bands \citep[see, e.g.,][]{Soszynski2008,Sicignano2024}, so applying PL/PW relations 
   derived solely from BL~Her and W~Vir stars to RV~Tau stars is expected to underestimate the 
   distance to the studied galaxy. To investigate this, we calculated Wesenheit magnitudes in the 
   $ri$~bands using reddened average magnitudes of M31 T2Ceps and reddening vectors from 
   \citet{Green2019}. Following \citet{Ngeow2022T2Cep}, we rejected six stars having period 
   uncertainty larger than one day. 
   Subsequently, we fitted our PW relation, derived using the parallax method for the combined 
   sample (see Table~\ref{tab:plr}), with fixed slope and allowing the ZP to vary as a~free 
   parameter (see Fig.~\ref{fig:comp_M31}).
   The resulting DM to M31 is $24.002 \pm 0.058$~mag. This value is significantly smaller than 
   the DM of \citet[][equals to $24.407 \pm 0.032$~mag]{Li2021}, calculated from classical 
   Cepheids using Hubble Space Telescope data, or the value of $24.46 \pm 0.10$~mag recommended 
   by \citet{deGrijsBono2014}. When fitting the same PW relation to W~Vir stars only (with a~fixed 
   slope and intercept as a~free parameter) we obtain $\mu = 24.253 \pm 0.071$~mag. Similarly,  
   applying the PW relation from Table~\ref{tab:plr}, specifically derived for W~Vir stars and 
   fitted to M31 W~Vir stars, yields $\mu = 24.255 \pm 0.108$~mag. Both these values are smaller 
   than the literature DM, revealing another potential problem, namely selection effect. 
   Fainter W~Vir stars in M31 may not have been detected by \citet{Kodric2018}, leading us to fit 
   the PW relations to only the brighter stars, which again shortens the inferred distance to M31.
   Another selection effect arises due to the very large halo size of M31 
   \citep[as reported by, e.g.,][]{Ibata2014}, which can cause stars lying on the far side of 
   the galaxy (farther away from us, and hence fainter) to remain undetected.
   Consequently, the distance to the M31 measured based on T2Ceps predominantly located in the 
   near side of its halo is likely underestimated. 
   Another potential difficulty is that \citet{Kodric2018} do not distinguish pW~Vir in their 
   sample of T2Ceps. However, the presence of such stars among brighter objects within this range 
   of periods is highly probable, as they are observed in other galaxies 
   \citep[see for example Fig.~$4$ from][]{Soszynski2018}. 
   We have inspected the light curves of the brightest variables classified as T2Ceps by 
   \citet{Kodric2018} with period less than $20$~days (i.e., in the range of W~Vir stars). All of 
   them have very similar light-curve shapes to pWVir stars with similar periods in the OGLE 
   catalogs of Magellanic Clouds variables. Therefore, this bias is certainly present in our 
   distance estimates to M31.
   
   The above comparison and application of our PL/PW relations with the literature are not 
   entirely satifactory. Nevertheless, because of the reasons discussed, this does not imply 
   that the relationships we calibrated in this work are incorrect. Instead, the comparison 
   shows the complexities of using T2Ceps to determine distances, particularly in optical bands.
   This consideration will be important for future studies employing Rubin-LSST data.


\subsubsection{Comparison of the PW relations for different types of pulsating stars}
\label{ssec:comp_all_var} 


   Figure~\ref{fig:pwr_all_var} presents a~comparison of PW relations (specifically for 
   $W^{gi}_{g}$) for Galactic classical pulsators. These include relations derived by 
   \citet[][for classical Cepheids pulsating in fundamental mode]{Narloch2023}, 
   \citet[][for RR~Lyrae stars of RRab and RRc types]{Narloch2024}, and those obtained in this 
   work for T2Ceps. 
   The plot closely resembles analogous optical/NIR relations for pulsating stars in the MW 
   \citep[e.g.][]{Ripepi2018,Ripepi2023,Soszynski2017,Soszynski2020} as well as in other galaxies 
   \citep[e.g.][]{Matsunaga2011,Soszynski2019}. This figure summarizes our work on the PL 
   relations of pulsating stars in the Sloan bands in our Galaxy using data from the LCO $40$~cm 
   telescopes, while also highlighting the potential for future research. 
   In particular, increasing the sample sizes of RR~Lyrae stars and T2Ceps would significantly 
   improve the precision and reliability of these relationships. 


\section{Summary} \label{sec:summary}


   In this work, we calibrated the PL and PW relations for MW T2Ceps of BL~Her and W~Vir types 
   separately, as well as for the combined sample, in the Sloan--Pan-STARRS $g_{P1}r_{P1}i_{P1}$ 
   bands and Wesenheit indices $W^{ri}_r$, $W^{gr}_r$, and $W^{gi}_g$. 
   Obtained coefficients are presented in Table~\ref{tab:plr}. 
   According to our knowledge, these are the first such relations for T2Ceps from the vicinity of 
   the Sun in these bands. 
   Scientific images were collected with the $40$~cm telescopes of the LCO Global Telescope 
   Network in the $g'r'i'$ filters. The instrumental magnitudes were calibrated to the 
   Sloan--Pan-STARRS version of the Sloan photometric system using the ATLAS-REFCAT2 catalog 
   \citep{Tonry2018}. The obtained light curves, presented in the Appendix~\ref{app:lc}, were 
   used to calculate intensity-averaged mean magnitudes, which are given in 
   Table~\ref{tab:t2cep}. These include values for $16$ BL~Her stars, $17$ W~Vir stars, as well 
   as three pW~Vir stars, two RV~Tau stars, and one first-overtone classical Cepheid.  
   We adopted reddening values for all our targets from the reddening map of \citet{SF2011}, 
   and the extinction coefficients were taken from \citet{Green2019}. 
   The absolute magnitudes of our objects were calculated using four methods: the inverse of the 
   Gaia DR3 parallaxes, the ABL method, and geometric and photogeometric distances from 
   \citet{BJ2021}. 
   Table~\ref{tab:plr} summarizes the results for all four methods, which can be used depending 
   on the needs.
   We have manually excluded some stars from the calibration of the PL/PW relations due to 
   reddening issues or likely incorrect Gaia DR3 parallax values (see Sect.~\ref{ssec:rej}). 
   
   We have discussed the influence of the ZP corrections by comparing the effects of applying 
   those from \citet{Groenewegen2021} instead of the ones provided by 
   \citet{LindegrenBastian2021}, as well as the impact of not introducing any corrections (see 
   ection~\ref{ssec:ZPinfluence}). 
   
   Finally, we compared our PL/PW relations for MW T2Ceps with those available in the literature 
   for the Sloan--Pan-STARRS photometric system, namely 
   \citet[][in globular clusters]{Ngeow2022T2Cep} and \citet[][for T2Ceps in M31]{Kodric2018}. 
   The comparison with \citet{Ngeow2022T2Cep} shows good agreement in the slope values; however, 
   there is a~significant difference in the ZPs (see Sect.~\ref{ssec:comp_with_lit}). The 
   distances to M31 calculated using the PW relations in the $ri$~bands do not reproduce the DM 
   derived from classical Cepheids \citep[from][]{Li2021}, most likely for reasons discussed in 
   Sect.~\ref{ssec:comp_with_lit}. 
   This disagreement does not necessarily indicate an error in our relations but rather potential 
   complications when using T2Ceps as distance indicators.


%
\begin{table*}
\caption{Determined coefficients of PL and PW relations for Galactic Type~II Cepheids.}             
\label{tab:plr}      
\centering          
\begin{tabular}{l|l|cccc|cccc}     
\hline\hline       
band & type & $a_{\lambda}$ & $b_{\lambda}$ & rms & N & 
$a_{\lambda}$ & $b_{\lambda}$ & rms & N  \\
\hline     
& & \multicolumn{4}{l}{(parallax):} & \multicolumn{4}{l}{(ABL method):} \\ 
\hline
\multirow{5}{*}{BL~Her} & $g$ & -1.231 $\pm$ 0.528 & -0.174 $\pm$ 0.077 & 0.24 & 14 & -1.137 $\pm$ 0.597 & -0.153 $\pm$ 0.087 & 0.24 & 14 \\ 
                        & $r$ & -1.430 $\pm$ 0.461 & -0.455 $\pm$ 0.068 & 0.21 & 14 & -1.323 $\pm$  0.534 & -0.436 $\pm$ 0.078 & 0.21 & 14 \\ 
                        & $i$ & -1.729 $\pm$ 0.442 & -0.513 $\pm$ 0.065 & 0.20 & 14 & -1.629 $\pm$ 0.531 & -0.495 $\pm$ 0.078 & 0.20 & 14 \\ 
\multirow{1}{*}{($\rm{log}P_0 = 0.3$)}  & $W^{ri}_r$ & -2.431 $\pm$ 0.483 & -0.653 $\pm$ 0.069 & 0.23 & 15 & -2.381 $\pm$ 0.617 & -0.636 $\pm$ 0.090 & 0.23 & 15 \\ 
  & $W^{gr}_r$ & -1.996 $\pm$ 0.428 & -1.270 $\pm$ 0.061 & 0.21 & 15 & -1.923 $\pm$ 0.504 & -1.255 $\pm$ 0.073 & 0.21 & 15 \\ 
  & $W^{gi}_g$ & -2.240 $\pm$ 0.383 & -0.923 $\pm$ 0.055 & 0.19 & 15 & -2.155 $\pm$ 0.457 & -0.909 $\pm$ 0.066 & 0.19 & 15 \\                         
\hline
\multirow{5}{*}{W~Vir} & $g$ & -1.144 $\pm$ 1.322 & -1.557 $\pm$ 0.139 & 0.39 & 10 & -1.251 $\pm$ 1.231 & -1.530 $\pm$ 0.138 & 0.40 & 10 \\ 
  & $r$ & -1.180 $\pm$ 1.170 & -2.017 $\pm$ 0.123 & 0.35 & 10 & -1.236 $\pm$ 1.104 & -1.995 $\pm$ 0.123 & 0.35 & 10 \\
  & $i$ & -1.297 $\pm$ 1.087 & -2.172 $\pm$ 0.114 & 0.32 & 10 & -1.322 $\pm$ 1.045 & -2.151 $\pm$ 0.117 & 0.32 & 10 \\ 
\multirow{1}{*}{($\rm{log}P_0 = 1.2$)}  & $W^{ri}_r$ & -1.635 $\pm$ 0.930 & -2.648 $\pm$ 0.096 & 0.28 & 11 & -1.540 $\pm$ 0.933 & -2.627 $\pm$ 0.103 & 0.28 & 11 \\ 
  & $W^{gr}_r$ & -1.214 $\pm$ 1.082 & -3.366 $\pm$ 0.111 & 0.33 & 11 & -1.123 $\pm$ 1.024 & -3.340 $\pm$ 0.111 & 0.33 & 11 \\ 
  & $W^{gi}_g$ & -1.450 $\pm$ 0.971 & -2.963 $\pm$ 0.100 & 0.29 & 11 & -1.357 $\pm$ 0.948 & -2.941 $\pm$ 0.104 & 0.30 & 11 \\ 
\hline
\multirow{5}{*}{BL~Her+W~Vir} & $g$ & -1.508 $\pm$ 0.135 & -0.804 $\pm$ 0.066 & 0.31 & 24 & -1.471 $\pm$ 0.155 & -0.777 $\pm$ 0.077 & 0.31 & 24 \\ 
  & $r$ & -1.709 $\pm$ 0.120 & -1.167 $\pm$ 0.058 & 0.27 & 24 & -1.673 $\pm$ 0.144 & -1.144 $\pm$ 0.072 & 0.28 & 24 \\
  & $i$ & -1.838 $\pm$ 0.112 & -1.262 $\pm$ 0.054 & 0.26 & 24 & -1.817 $\pm$ 0.143 & -1.243 $\pm$ 0.072 & 0.26 & 24 \\ 
\multirow{1}{*}{($\rm{log}P_0 = 0.7$)}  & $W^{ri}_r$ & -2.251 $\pm$ 0.108 & -1.543 $\pm$ 0.051 & 0.25 & 26 & -2.275 $\pm$ 0.176 & -1.536 $\pm$ 0.088 & 0.25 & 26 \\ 
  & $W^{gr}_r$ & -2.315 $\pm$ 0.115 & -2.229 $\pm$ 0.055 & 0.27 & 26 & -2.276 $\pm$ 0.156 & -2.204 $\pm$ 0.078 & 0.27 & 26 \\ 
  & $W^{gi}_g$ & -2.279 $\pm$ 0.101 & -1.844 $\pm$ 0.048 & 0.24 & 26 & -2.270 $\pm$ 0.138 & -1.830 $\pm$ 0.069 & 0.24 & 26 \\ 
\hline 
& & \multicolumn{4}{l}{(geometric):} & \multicolumn{4}{l}{(photogeometric):} \\ 
\hline
\multirow{5}{*}{BL~Her} & $g$ & -1.218 $\pm$ 0.511 & -0.182 $\pm$ 0.075 & 0.24 & 14 & -1.149 $\pm$ 0.557 & -0.154 $\pm$ 0.082 & 0.26 & 14 \\ 
  & $r$ & -1.417 $\pm$ 0.441 & -0.463 $\pm$ 0.065 & 0.20 & 14 & -1.348 $\pm$ 0.493 & -0.435 $\pm$ 0.072 & 0.23 & 14 \\
  & $i$ & -1.716 $\pm$ 0.422 & -0.521 $\pm$ 0.062 & 0.19 & 14 & -1.647 $\pm$ 0.471 & -0.493 $\pm$ 0.069 & 0.22 & 14 \\ 
\multirow{1}{*}{($\rm{log}P_0 = 0.3$)}  & $W^{ri}_r$ & -2.415 $\pm$ 0.471 & -0.660 $\pm$ 0.068 & 0.23 & 15 & -2.359 $\pm$ 0.492 & -0.634 $\pm$ 0.070 & 0.24 & 15 \\ 
  & $W^{gr}_r$ & -1.981 $\pm$ 0.409 & -1.278 $\pm$ 0.059 & 0.20 & 15 & -1.924 $\pm$ 0.454 & -1.252 $\pm$ 0.065 & 0.22 & 15 \\ 
  & $W^{gi}_g$ & -2.225 $\pm$ 0.365 & -0.931 $\pm$ 0.052 & 0.18 & 15 & -2.168 $\pm$ 0.401 & -0.905 $\pm$ 0.058 & 0.19 & 15 \\                 
\hline
\multirow{5}{*}{W~Vir} & $g$ & -1.082 $\pm$ 1.347 & -1.579 $\pm$ 0.142 & 0.40 & 10 & -1.107 $\pm$ 1.298 & -1.515 $\pm$ 0.136 & 0.39 & 10 \\ 
  & $r$ & -1.118 $\pm$ 1.199 & -2.039 $\pm$ 0.126 & 0.36 & 10 & -1.143 $\pm$ 1.142 & -1.975 $\pm$ 0.120 & 0.34 & 10 \\
  & $i$ & -1.235 $\pm$ 1.112 & -2.194 $\pm$ 0.117 & 0.33 & 10 & -1.260 $\pm$ 1.068 & -2.130 $\pm$ 0.112 & 0.32 & 10 \\ 
\multirow{1}{*}{($\rm{log}P_0 = 1.2$)}  & $W^{ri}_r$ & -1.588 $\pm$ 0.932 & -2.667 $\pm$ 0.096 & 0.28 & 11 & -1.585 $\pm$ 0.946 & -2.608 $\pm$ 0.097 & 0.29 & 11 \\ 
  & $W^{gr}_r$ & -1.167 $\pm$ 1.107 & -3.385 $\pm$ 0.114 & 0.34 & 11 & -1.165 $\pm$ 1.059 & -3.326 $\pm$ 0.109 & 0.32 & 11 \\ 
  & $W^{gi}_g$ & -1.403 $\pm$ 0.985 & -2.981 $\pm$ 0.101 & 0.30 & 11 & -1.401 $\pm$ 0.969 & -2.923 $\pm$ 0.100 & 0.29 & 11 \\ 
\hline
\multirow{5}{*}{BL~Her+W~Vir} & $g$ & -1.522 $\pm$ 0.136 & -0.820 $\pm$ 0.066 & 0.31 & 24 & -1.478 $\pm$ 0.137 & -0.776 $\pm$ 0.066 & 0.31 & 24 \\ 
  & $r$ & -1.722 $\pm$ 0.121 & -1.183 $\pm$ 0.058 & 0.28 & 24 & -1.678 $\pm$ 0.121 & -1.139 $\pm$ 0.059 & 0.28 & 24 \\
  & $i$ & -1.851 $\pm$ 0.112 & -1.278 $\pm$ 0.055 & 0.26 & 24 & -1.807 $\pm$ 0.114 & -1.234 $\pm$ 0.055 & 0.26 & 24 \\ 
\multirow{1}{*}{($\rm{log}P_0 = 0.7$)}  & $W^{ri}_r$ & -2.262 $\pm$ 0.107 & -1.558 $\pm$ 0.051 & 0.25 & 26 & -2.222 $\pm$ 0.110 & -1.517 $\pm$ 0.052 & 0.26 & 26 \\ 
  & $W^{gr}_r$ & -2.326 $\pm$ 0.116 & -2.244 $\pm$ 0.055 & 0.27 & 26 & -2.286 $\pm$ 0.116 & -2.203 $\pm$ 0.055 & 0.27 & 26 \\ 
  & $W^{gi}_g$ & -2.290 $\pm$ 0.101 & -1.858 $\pm$ 0.048 & 0.24 & 26 & -2.250 $\pm$ 0.103 & -1.818 $\pm$ 0.049 & 0.24 & 26 \\ 
\hline
\end{tabular}
\tablefoot{
    type: variability class;  
    band: the Pan-STARRS $g_{P1}r_{P1}i_{P1}$ bands; 
    $a_{\lambda}$: slope of the fit; 
    $b_{\lambda}$: zero point of the fit; 
    rms: a~root mean square of derived relations; 
    N: number of stars used for fitting.}
\end{table*}

%
   \begin{figure*}
   \centering
   \includegraphics[width=0.9\hsize]{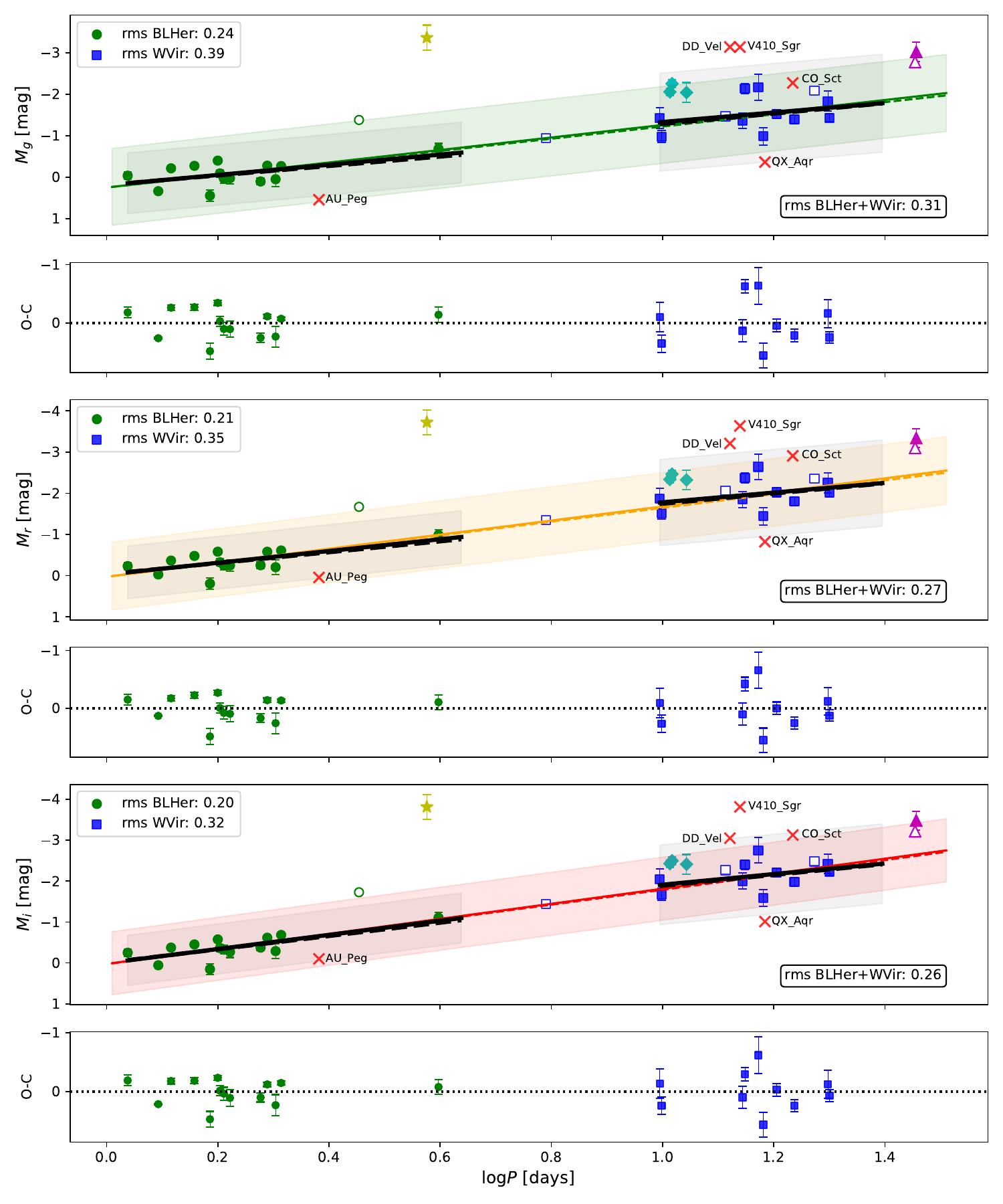}
   \caption{PL relations for T2Ceps based on the reddening values from \citet{SF2011} 
            and reddening vectors ($R_{\lambda}$) from \citet{Green2019}. 
            Filled circles and squares: BL~Her and W~Vir type stars adopted for derivation of 
            PL relations; 
            open marks: stars with $RUWE>1.4$; 
            red crosses: stars manually rejected from the fit (see 
            Sect.~\ref{sec:relations} for explanation); 
            yellow star-symbol: first-overtone Type~I Cepheid; 
            cyan diamonds: pW~Vir type stars; 
            magenta triangles: RV~Tau type stars;
            solid lines: the fit to Eq.~(\ref{eq:plr}) for BL~Her and W~Vir stars 
            separately, as well as combined sample; 
            dashed line: analogous fit to Eq.~(\ref{eq:abl_plr}); 
            shaded areas: $\pm 3$rms for BL~Her and W~Vir stars and combined sample. 
            The pivot logarithms of $\rm{log}P_0=0.3;\,1.2;\,0.7$ were used to fit the BL~Her, 
            W~Vir and combined sample, respectively. 
            }
      \label{fig:plr_blher+wvir}
    \end{figure*}
%

%
   \begin{figure*}
   \centering
   \includegraphics[width=0.9\hsize]{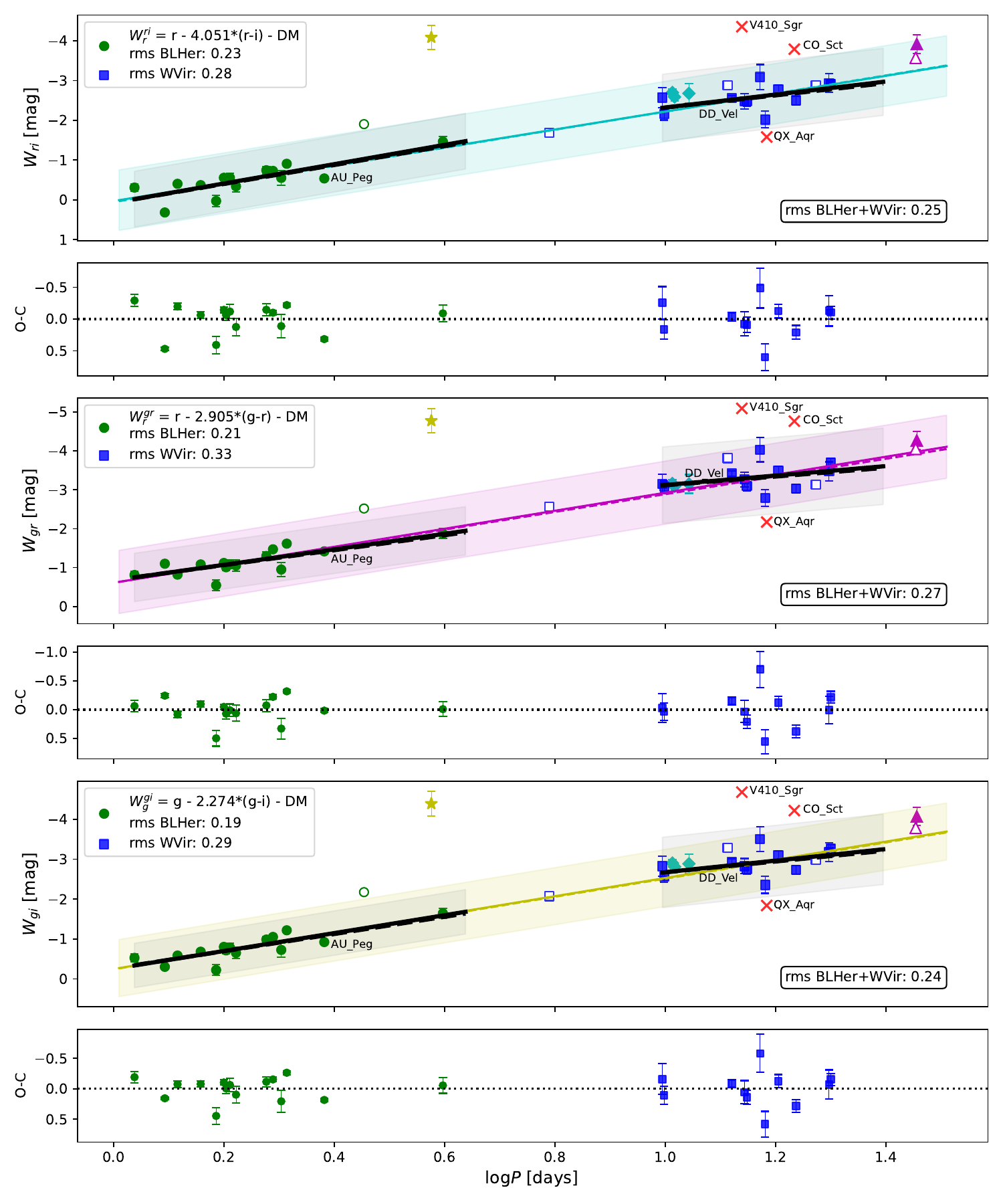}
   \caption{PW relations for T2Ceps based on the reddening vector ($R_{\lambda}$) from 
            \citet{Green2019}.  
            Filled circles and squares: BL~Her and W~Vir type stars adopted for derivation of 
            PW relations; 
            open marks: stars with $RUWE>1.4$; 
            red crosses: stars manually rejected from the fit (see 
            Sect.~\ref{sec:relations} for explanation); 
            yellow star-symbol: first-overtone Type~I Cepheid; 
            cyan diamonds: pW~Vir type stars; 
            magenta triangles: RV~Tau type stars;
            solid lines: the fit to Eq.~(\ref{eq:pwr}) for BL~Her and W~Vir stars 
            separately, as well as combined sample; 
            dashed line: analogous fit to Eq.~(\ref{eq:abl_pwr}); 
            shaded areas: $\pm 3$rms for BL~Her and W~Vir stars and combined sample. 
            The pivot logarithms of $\rm{log}P_0=0.3;\,1.2;\,0.7$ were used to fit the BL~Her, 
            W~Vir and combined sample, respectively. 
            }
      \label{fig:pwr_blher+wvir}
    \end{figure*}
%

%
   \begin{figure*}
   \centering
   \includegraphics[width=0.612\hsize]{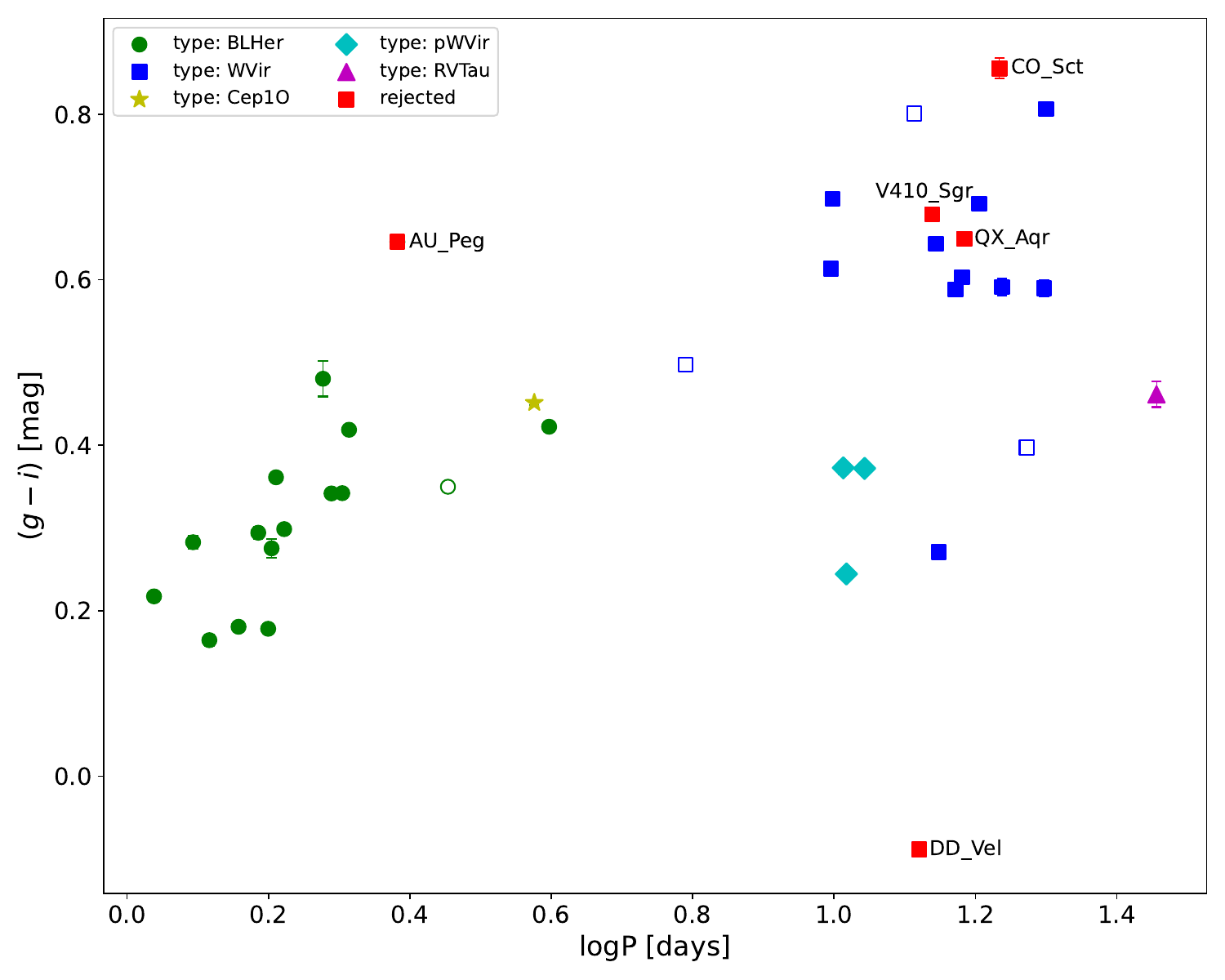}
   \includegraphics[width=0.368\hsize]{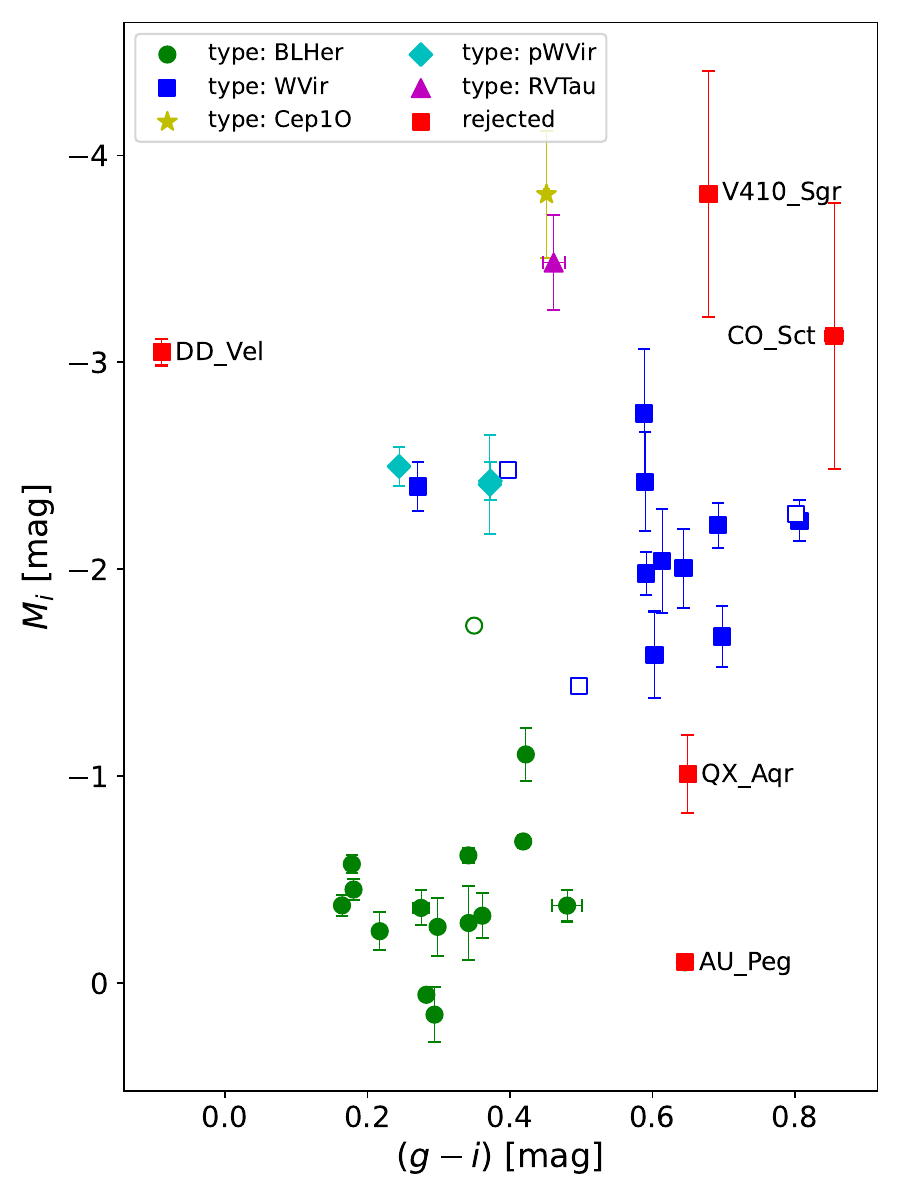}
   \caption{Color index $(g-i)$ of T2Ceps from the MW as a~function of the logarithm of 
            pulsational period (left panel) and color-magnitude diagram from those stars 
            (right panel).
            }
      \label{fig:cmd}
    \end{figure*}
%

%
   \begin{figure*}
   \centering
   \includegraphics[width=0.9\hsize]{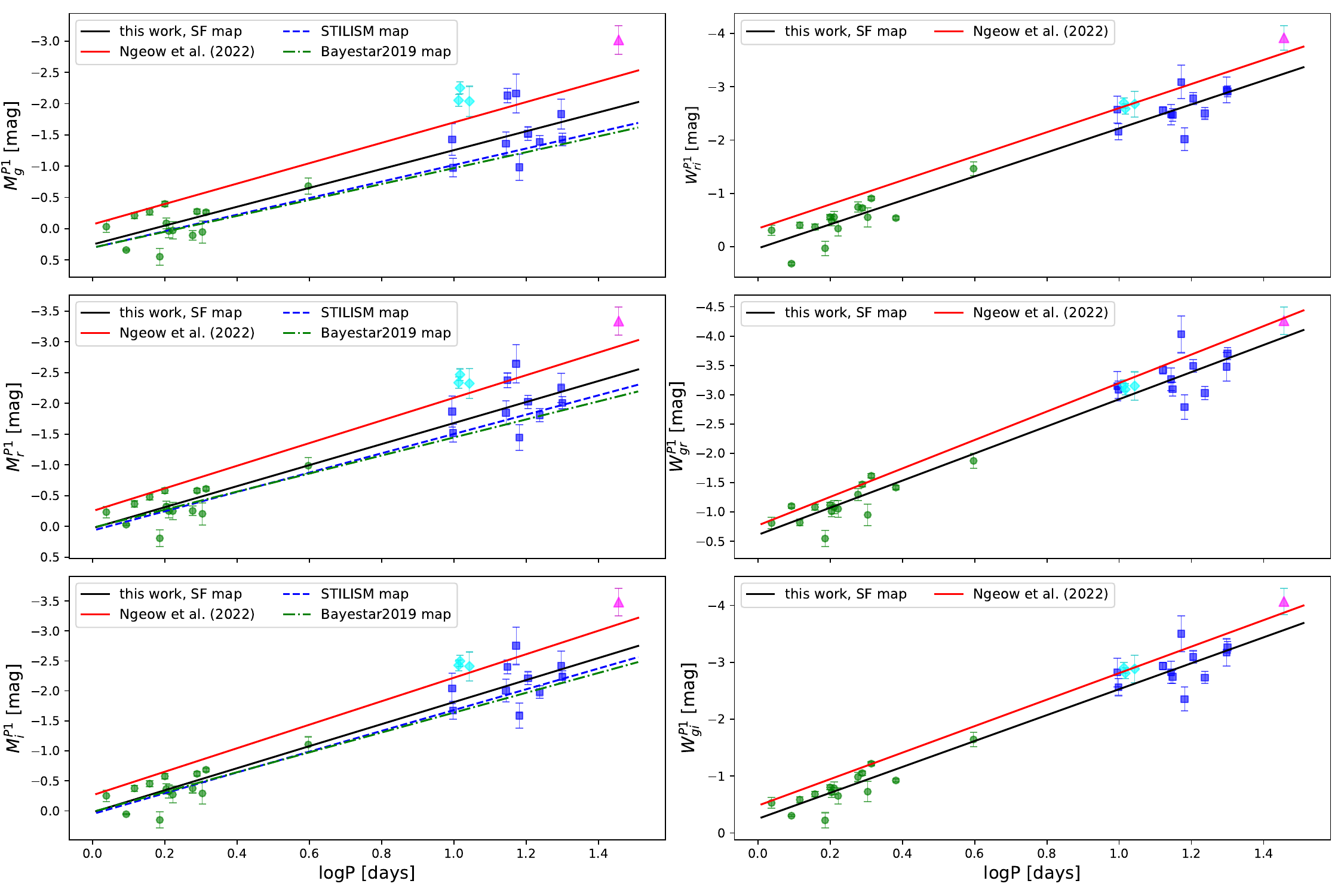}
   \caption{Comparison of the PL/PW relations derived in this work for the combined sample of 
            BL~Her and W~Vir stars (black lines) with the PL/PW relations determined by 
            \citet[][see their Tables~$3$ and $4$]{Ngeow2022T2Cep} for T2Ceps from MW globular 
            clusters (red lines). Blue dashed and green dashdot lines in the left panels mark 
            PL/PW relations derived using STILISM and Bayestar2019 reddening maps, respectively.
            }
      \label{fig:comp_Ngeow2022}
    \end{figure*}
%

%
   \begin{figure*}
   \centering
   \includegraphics[width=0.9\hsize]{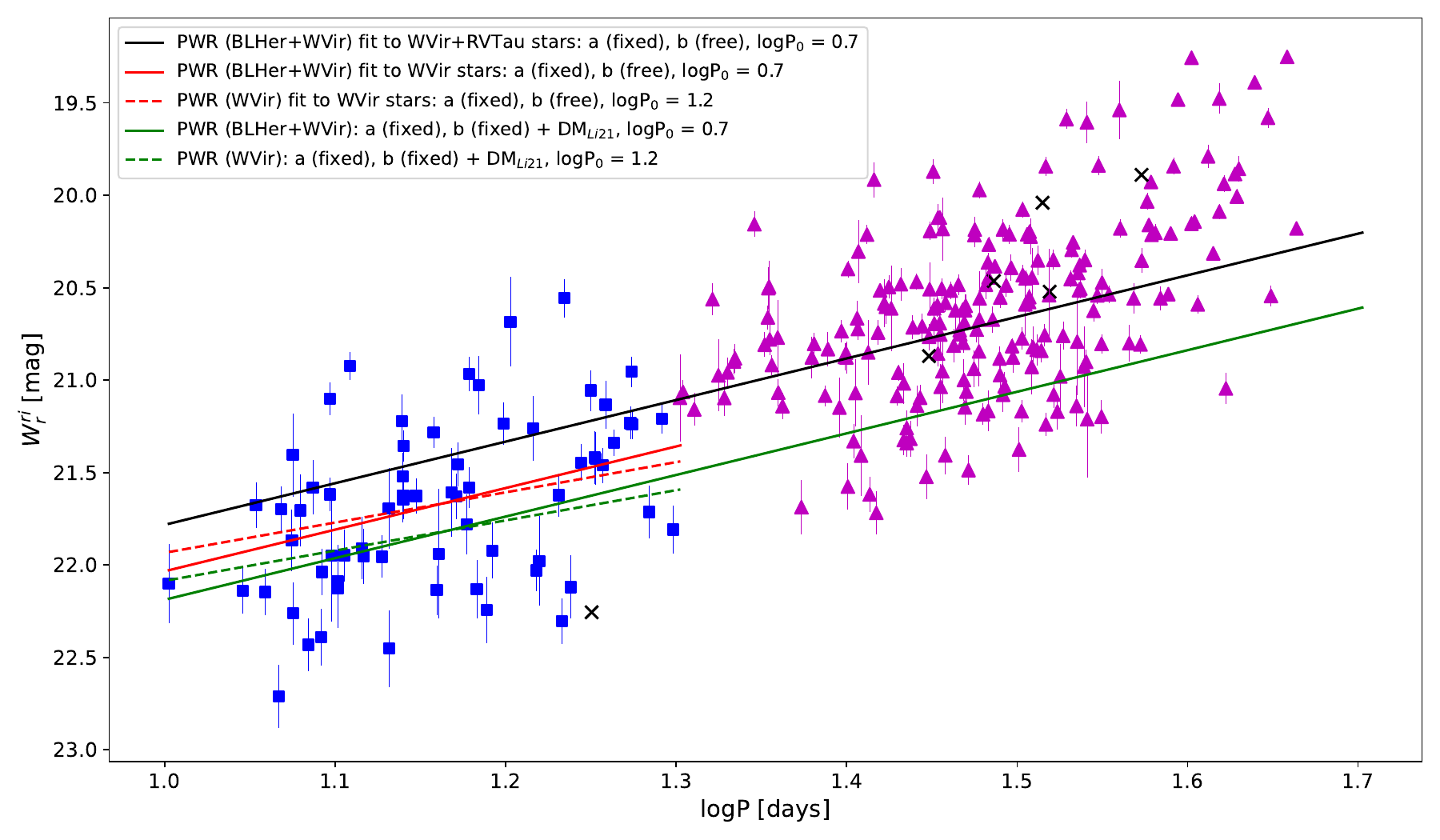}
   \caption{PW relation for T2Ceps from \citet{Kodric2018} calculated using the reddening vectors 
            from \citet{Green2019}. Blue squares mark W~Vir type stars; 
            magenta triangles mark RV~Tau type stars; and black crosses mark stars with period 
            uncertainty larger than one day. 
            Lines mark PW relations as defined by Eq.~\ref{eq:pwr} for the combined 
            sample of T2Ceps (solid lines) and solely W~Vir stars (dahsed lines). 
            Green lines mark PW relations derived in this work, shifted by the DM from 
            \citet[][$\mu = 24.407$~mag]{Li2021}. 
            }
      \label{fig:comp_M31}
    \end{figure*}
%

%
   \begin{figure*}
   \centering
   \includegraphics[width=0.9\hsize]{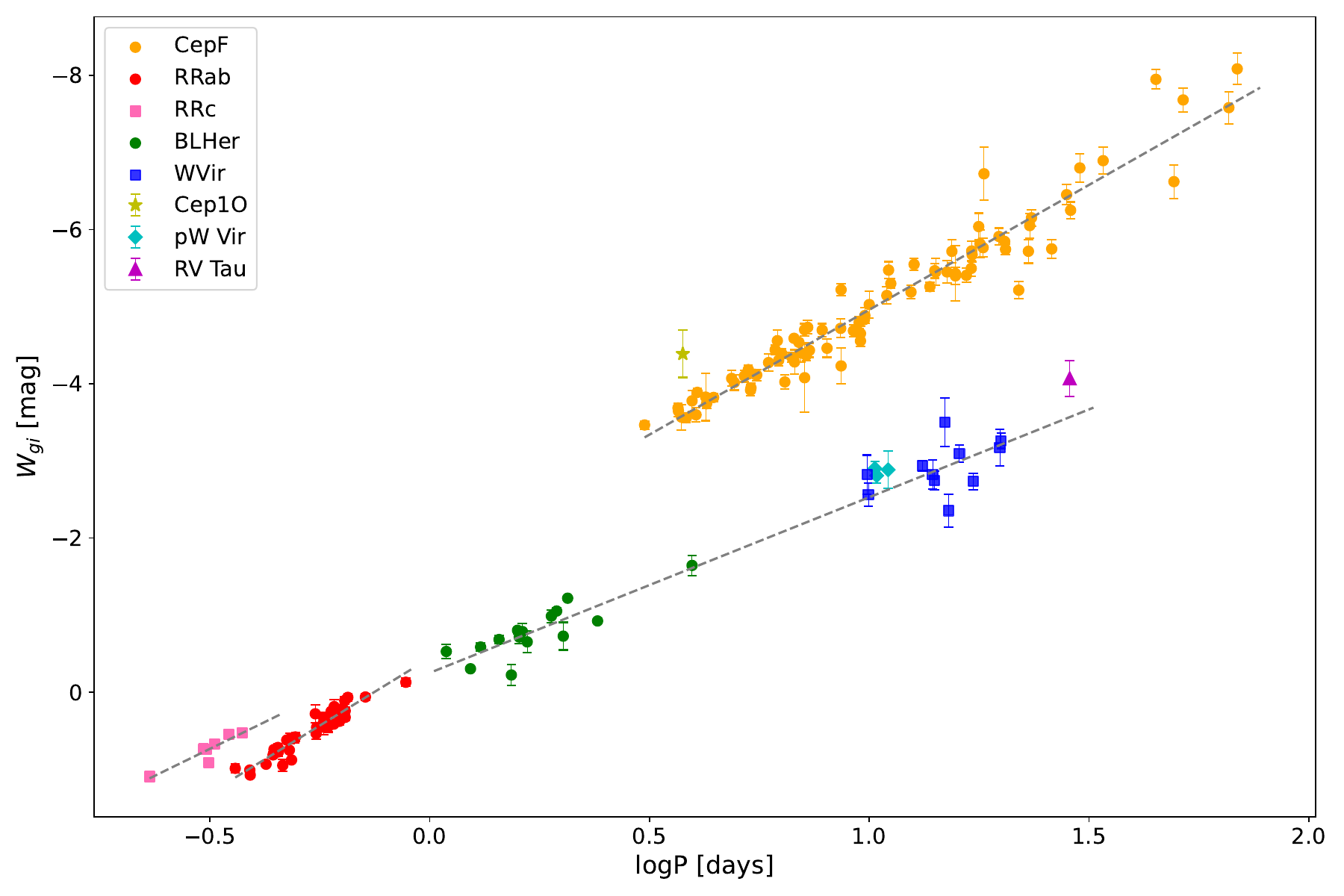}
   \caption{Compilation of the PW relations for pulsating variables from the MW in the Sloan 
            bands. Gray, dashed lines are the PW relations derived for each type of variables 
            in \citet[][for classical Cepheids]{Narloch2023}, 
            \citet[][for RR~Lyrae stars]{Narloch2024} and this work (T2Ceps). 
            }
      \label{fig:pwr_all_var}
    \end{figure*}
%

\begin{acknowledgements}
      We thank the anonymous referee for valuable comments which improved this paper.
      The research leading to these results has received funding from the European 
      Research Council (ERC) under the European Union’s Horizon 2020 research and 
      innovation program (grant agreement No. 695099). We also acknowledge support 
      from the National Science Center, Poland grants MAESTRO UMO-2017/26/A/ST9/00446, 
      BEETHOVEN UMO-2018/31/G/ST9/03050 and DIR/WK/2018/09 grants of the Polish Ministry 
      of Science and Higher Education. We also acknowledge financial support from the 
      European Research Council (ERC) under the European Union’s Horizon 2020 
      research and innovation program (grant agreement No. 951549 - UniverScale), 
      as well as the Polish Ministry of Science and Higher Education grant agreement 
      2024/WK/02; and National Science Center, Poland, Sonata BIS project 2018/30/E/ST9/00598. 
      We gratefully acknowledge financial support for this work from the BASAL Centro de 
      Astrofisica y Tecnologias Afines (CATA) AFB-170002 and the Millenium Institute 
      of Astrophysics (MAS) of the Iniciativa Cientifica Milenio del Ministerio de 
      Economia, Fomento y Turismo de Chile, project IC120009. 
      W.G. also gratefully acknowledges support from the ANID BASAL project ACE210002. 
      P.W. gratefully acknowledges financial support from the Polish National Science 
      Center grant PRELUDIUM 2018/31/N/ST9/02742. 
      This work has made use of data from the European Space Agency (ESA) mission Gaia 
      (\url{https://www.cosmos.esa.int/gaia}), processed by the Gaia Data Processing and 
      Analysis Consortium (DPAC, \url{https://www.cosmos.esa.int/web/gaia/dpac/consortium}). 
      Funding for the DPAC has been provided by national institutions, in particular the 
      institutions participating in the Gaia Multilateral Agreement. 
      
      Facilities: LCOGT (0.4m).
      
      Software used in this work: {\tt gaiadr3\_zero-point} \citep{LindegrenBastian2021}, 
          IRAF \citep{Tody1986,Tody1993}, 
          DAOPHOT \citep{Stetson1987}, 
          Astropy \citep{Astropy2013},  
          Sklearn \citep{Pedregosa2011}
          NumPy \citep{Numpy1,Numpy2}, 
          SciPy \citep{Virtanen2020}, 
          Matplotlib \citep{Hunter2007}.
\end{acknowledgements}

%
%

\bibliographystyle{aa} 
\bibliography{ms_arXiv} 

%
%
%
%
%
%
%
%
%

%

\begin{appendix} 


\onecolumn
\section{Fourier analysis of Type~II Cepheids.} 
\label{app:fp}

We analyzed $g_{P1}$-band light curves of stars from our sample with Fourier analysis in order 
to confirm their classification into a~certain group of pulsating stars. The derived parameters, 
including relative amplitudes and Fourier phases of harmonics \citep{Simon1981} are listed in 
Table~\ref{tab:fp} and shown in Fig.~\ref{fig:fp}. 
We compared obtained diagrams with analogous diagrams from \citet{Soszynski2008} and 
\citet{Soszynski2018} for the Magellanic Clouds, as well as \citet{Soszynski2017} and 
\citet{Soszynski2020} for Galacitc Disk and bulge for the reference.
In Fig.~\ref{fig:fp}, BL~Her type 
stars are marked with circles, while W~Vir type stars are marked with squares. 
Different types of pulsating stars are represented with different colors. 
The pW~Vir type stars (marked with blue squares) clearly deviate from the rest of the T2Ceps, 
particularly in the $\phi_{21}$ parameter. Stars manually rejected from the fitting of the PL/PW 
relations (except DD~Vel and AU~Peg, which were rejected from the PL relation only) are marked 
with red squares. Their Fourier parameters seem to confirm their classification as W~Vir type 
stars, which suggests that their deviation from the relation is not due to misclassification but 
rather due to the reasons mentioned in the Sect.~\ref{sec:relations}. 

\begin{table*}[h!]
\scriptsize
\caption{\label{tab:fp} Fourier parameters determined for stars from our sample based on 
$g_{P1}$-band light curves.}
\begin{tabular}{lcccccccc}
\hline\hline 
Name & P$_{lit}$ & N & m$_{0}$ & $A_{1}$ & $R_{21}$ & $R_{31}$ & $\phi_{21}$ & $\phi_{31}$ \\
     &  (days)   &   & (mag)   &         &          &          &             &             \\ 
\hline
\multicolumn{9}{l}{BL~Her type stars} \\ 
\hline
AU~Peg & 2.4121 & 2 & 9.5973 & 0.18121 $\pm$ 0.00781 & 0.14606 $\pm$ 0.04590 & 3.37202 $\pm$ 0.31702 & - & - \\ 
BL~Her & 1.3074 & 7 & 10.3761 & 0.40218 $\pm$ 0.00788 & 0.33692 $\pm$ 0.02130 & 3.11465 $\pm$ 0.07262 & 0.28382 $\pm$ 0.02103 & 0.06807 $\pm$ 0.08946 \\
BX~Del & 1.0918 & 6 & 12.4256 & 0.37086 $\pm$ 0.00712 & 0.34898 $\pm$ 0.02117 & 3.04093 $\pm$ 0.06051 & 0.19151 $\pm$ 0.01898 & 5.83051 $\pm$ 0.12228 \\
FM~Del & 3.9540 & 4 & 12.6745 & 0.37267 $\pm$ 0.00590 & 0.27642 $\pm$ 0.01579 & 3.55653 $\pm$ 0.06578 & 0.08242 $\pm$ 0.01592 & 0.66571 $\pm$ 0.19137 \\ 
FY~Aqr & 1.0229 & 7 & 12.5612 & 0.34999 $\pm$ 0.00393 & 0.40337 $\pm$ 0.01145 & 2.32183 $\pm$ 0.03348 & 0.27927 $\pm$ 0.01162 & 4.62980 $\pm$ 0.05030 \\ 
NW~Lyr & 1.6012 & 7 & 12.7135 & 0.44643 $\pm$ 0.01154 & 0.22228 $\pm$ 0.02516 & 3.10687 $\pm$ 0.15272 & 0.19063 $\pm$ 0.02395 & 4.83484 $\pm$ 0.18188 \\ 
RT~TrA & 1.9461 & 7 & 10.0708 & 0.39545 $\pm$ 0.00392 & 0.11675 $\pm$ 0.01045 & 3.22774 $\pm$ 0.08001 & 0.08033 $\pm$ 0.00986 & 4.20607 $\pm$ 0.12110 \\ 
SW~Tau & 1.5835 & 7 & 10.0458 & 0.41307 $\pm$ 0.00557 & 0.33374 $\pm$ 0.01370 & 2.71924 $\pm$ 0.05668 & 0.07562 $\pm$ 0.01410 & 0.23295 $\pm$ 0.17615 \\ 
V2022~Sgr & 1.5332 & 6 & 13.8764 & 0.39378 $\pm$ 0.00962 & 0.26345 $\pm$ 0.02745 & 2.90095 $\pm$ 0.10980 & 0.20519 $\pm$ 0.02526 & 4.02560 $\pm$ 0.17115 \\ 
V439~Oph & 1.8930 & 3$^{a}$ & 12.5573 & 0.39932 $\pm$ 0.07045 & 0.11548 $\pm$ 0.13613 & 5.75253 $\pm$ 0.95450 & 0.24045 $\pm$ 0.08430 & 3.97940 $\pm$ 0.41494 \\ 
V465~Oph & 2.8438 & 6 & 13.8687 & 0.57397 $\pm$ 0.01509 & 0.37541 $\pm$ 0.02652 & 3.30695 $\pm$ 0.08630 & 0.16710 $\pm$ 0.02612 & 6.26959 $\pm$ 0.18152 \\ 
V477~Oph & 2.0157 & 8 & 14.1192 & 0.40900 $\pm$ 0.00497 & 0.22662 $\pm$ 0.01339 & 3.18794 $\pm$ 0.07489 & 0.10586 $\pm$ 0.01298 & 5.52752 $\pm$ 0.13348 \\ 
V553~Cen & 2.0606 & 4 & 8.7331 & 0.29545 $\pm$ 0.00400 & 0.06802 $\pm$ 0.01387 & 1.87116 $\pm$ 0.20383 & 0.05469 $\pm$ 0.01370 & 3.59067 $\pm$ 0.25786 \\ 
V606~Pup & 1.4378 & 7 & 11.5367 & 0.27673 $\pm$ 0.00239 & 0.22220 $\pm$ 0.00878 & 3.60418 $\pm$ 0.04202 & 0.03067 $\pm$ 0.00852 & 5.79016 $\pm$ 0.28527 \\ 
V971~Aql & 1.6245 & 9 & 12.2697 & 0.37793 $\pm$ 0.00383 & 0.24337 $\pm$ 0.01019 & 3.05286 $\pm$ 0.04617 & 0.18968 $\pm$ 0.01051 & 4.57981 $\pm$ 0.05860 \\
VY~Pyx & 1.2400 & 2 & 7.5093 & 0.13675 $\pm$ 0.00839 & 0.12769 $\pm$ 0.06039 & 2.47041 $\pm$ 0.49393 & - & - \\ 
VZ~Aql & 1.6683 & 9 & 13.9701 & 0.45369 $\pm$ 0.00600 & 0.23149 $\pm$ 0.01326 & 3.29850 $\pm$0.06688 & 0.14200 $\pm$ 0.01274 & 4.79516 $\pm$ 0.10396 \\ 
\hline
\multicolumn{9}{l}{W~Vir type stars} \\ 
\hline
AL~Lyr & 12.9780 & 3 & 12.5396 & 0.40749 $\pm$ 0.00707 & 0.16609 $\pm$ 0.01808 & 4.59975 $\pm$ 0.11125 & 0.12770 $\pm$ 0.01839 & 3.83633 $\pm$ 0.14250 \\ 
AX~Tel & 9.8960 & 6 & 12.9652 & 0.37053 $\pm$ 0.00586 & 0.12727 $\pm$ 0.01540 & 3.56875 $\pm$ 0.12545 & 0.11518 $\pm$ 0.01604 & 3.50098 $\pm$ 0.13566 \\ 
CO~Pup & 16.0420 & 4 & 11.2715 & 0.54335 $\pm$ 0.00626 & 0.13224 $\pm$ 0.01160 & 5.28422 $\pm$ 0.09572 & 0.11707 $\pm$ 0.01193 & 4.47432 $\pm$ 0.10804 \\ 
CO~Sct & 17.1380 & 5$^{*}$ & 14.9487 & 0.74436 $\pm$ 0.01157 & 0.26533 $\pm$ 0.01451 & 4.14223 $\pm$ 0.05920 & 0.09209 $\pm$ 0.01416 & 6.12444 $\pm$ 0.14889 \\ 
DD~Vel & 13.1980 & 3 & 13.2093 & 0.38951 $\pm$ 0.00694 & 0.19006 $\pm$ 0.01986 & 5.35068 $\pm$ 0.09608 & 0.13737 $\pm$ 0.01792 & 4.08525 $\pm$ 0.14425 \\ 
FI~Sct & 14.8512 & 9 & 15.0368 & 0.67381 $\pm$ 0.01058 & 0.36235 $\pm$ 0.01560 & 4.18713 $\pm$ 0.05124 & 0.09465 $\pm$ 0.01522 & 5.66475 $\pm$ 0.17600 \\ 
HQ~Car & 14.0670 & 6 & 12.7951 & 0.56364 $\pm$ 0.00747 & 0.46169 $\pm$ 0.01540 & 2.95969 $\pm$ 0.03951 & 0.06685 $\pm$ 0.01342 & 5.30420 $\pm$ 0.20730 \\ 
HQ~Cen &  9.9585 & 4 & 12.3068 & 0.26072 $\pm$ 0.00628 & 0.22522 $\pm$ 0.02311 & 4.36174 $\pm$ 0.11353 & 0.04332 $\pm$ 0.02362 & 2.47776 $\pm$ 0.53241 \\ 
MR~Ara & 19.8140 & 3$^{*}$ & 11.7940 & 0.61366 $\pm$ 0.01187 & 0.16075 $\pm$ 0.01868 & 4.09238 $\pm$ 0.11959 & 0.05155 $\pm$ 0.01773 & 6.28234 $\pm$ 0.36489 \\
QX~Aqr & 15.2850 & 3 & 12.6038 & 0.49646 $\pm$ 0.00682 & 0.13280 $\pm$ 0.01436 & 5.75650 $\pm$ 0.11662 & 0.11748 $\pm$ 0.01481 & 4.92995 $\pm$ 0.12576 \\ 
RS~Pav & 19.9450 & 6$^{*}$ & 11.0343 & 0.70558 $\pm$ 0.00737 & 0.16996 $\pm$ 0.01080 & 4.34805 $\pm$ 0.06641 & 0.06142 $\pm$ 0.01061 & 3.76527 $\pm$ 0.17913 \\ 
ST~Pup & 18.7300 & 5 & 10.2985 & 0.58049 $\pm$ 0.02032 & 0.45838 $\pm$ 0.03867 & 3.42677 $\pm$ 0.10431 & 0.23706 $\pm$ 0.03597 & 6.18184 $\pm$ 0.18330 \\
TX~Del & 6.1659 & 3 & 9.5262 & 0.33226 $\pm$ 0.01143 & 0.24790 $\pm$ 0.03748 & 3.63268 $\pm$ 0.14691 & 0.06392 $\pm$ 0.03374 & 0.59673 $\pm$ 0.55149 \\
V410~Sgr & 13.7753 & 7 & 13.0645 & 0.51930 $\pm$ 0.00647 & 0.26158 $\pm$ 0.01314 & 4.27865 $\pm$ 0.05453 & 0.14703 $\pm$ 0.01257 & 5.15200 $\pm$ 0.09111 \\ 
V741~Sgr & 15.1682 & 8 & 13.2087 & 0.52404 $\pm$ 0.00787 & 0.17234 $\pm$ 0.01500 & 4.45275 $\pm$ 0.08832 & 0.10196 $\pm$ 0.01509 & 5.45217 $\pm$ 0.15123 \\
VZ~Tau & 13.9430 & 7 & 13.1807 & 0.61006 $\pm$ 0.00670 & 0.19004 $\pm$ 0.01095 & 4.69842 $\pm$ 0.05947 & 0.14457 $\pm$ 0.01100 & 4.62815 $\pm$ 0.08380 \\
W~Vir & 17.2736 & 5 & 10.2988 & 0.63390 $\pm$ 0.01180 & 0.22385 $\pm$ 0.01954 & 4.39178 $\pm$ 0.09238 & 0.09530 $\pm$ 0.01892 & 6.11618 $\pm$ 0.20189 \\ 
\hline
\multicolumn{9}{l}{other variable types} \\ 
\hline
V572~Aql & 3.7673 & 3 & 11.4682 & 0.23644 $\pm$ 0.00323 & 0.07118 $\pm$ 0.01411 & 1.80022 $\pm$ 0.19339 & 0.02846 $\pm$ 0.01409 & 2.47723 $\pm$ 0.48597 \\ 
AL~Vir & 10.3065 & 5 & 9.7950 & 0.46084 $\pm$ 0.01011 & 0.21292 $\pm$ 0.02148 & 1.85852 $\pm$ 0.09527 & 0.04200 $\pm$ 0.02346 & 4.26210 $\pm$ 0.39704 \\ 
AP~Her & 10.4110 & 3 & 11.1666 & 0.41771 $\pm$ 0.00568 & 0.22284 $\pm$ 0.01455 & 2.25886 $\pm$ 0.06566 & 0.09564 $\pm$ 0.01385 & 4.61761 $\pm$ 0.14454 \\
BH~Oph & 11.0480 & 4 & 12.3382 & 0.45976 $\pm$ 0.00617 & 0.29141 $\pm$ 0.01427 & 2.27396 $\pm$ 0.05548 & 0.06927 $\pm$ 0.01381 & 5.22771 $\pm$ 0.19984 \\ 
TW~Cap & 28.5850 & 6$^{*}$ & 10.8641 & 0.54425 $\pm$ 0.00767 & 0.49126 $\pm$ 0.01588 & 3.47829 $\pm$ 0.04115 & 0.25416 $\pm$ 0.01489 & 6.08272 $\pm$ 0.06972 \\
V1711~Sgr & 28.4600 & 1$^{b}$ & - & - & - & - & - & - \\
\hline
\multicolumn{9}{@{}l@{}}{\parbox{0.96\textwidth}{
     Name: name of the Type~II Cepheid; 
     P$_{lit}$: period given by AAVSO (as in Table~\ref{tab:t2cep}); 
     m$_{0}$: mean magnitude (calculated directly from magnitudes, without going through fluxes); 
     N: order of the Fourier fit used; 
     $A_{1}$, $R_{21}$, $R_{31}$, $\phi_{21}$, $\phi_{31}$: Fourier fit parameters; 
     err: error. \\ 
     $^{*}$ Parameters calculated for doubled period. \\
     $^{a}$ Because of the incomplete light curve, especially in the minimum of the brightness, 
     Fourier parameters are unreliable for this star. \\ 
     $^{b}$ Unstable light curve, covering specific, variable cycles. 
}}
\end{tabular}
\end{table*}

%
   \begin{figure*}[h!]
   \centering
   \includegraphics[width=0.9\hsize]{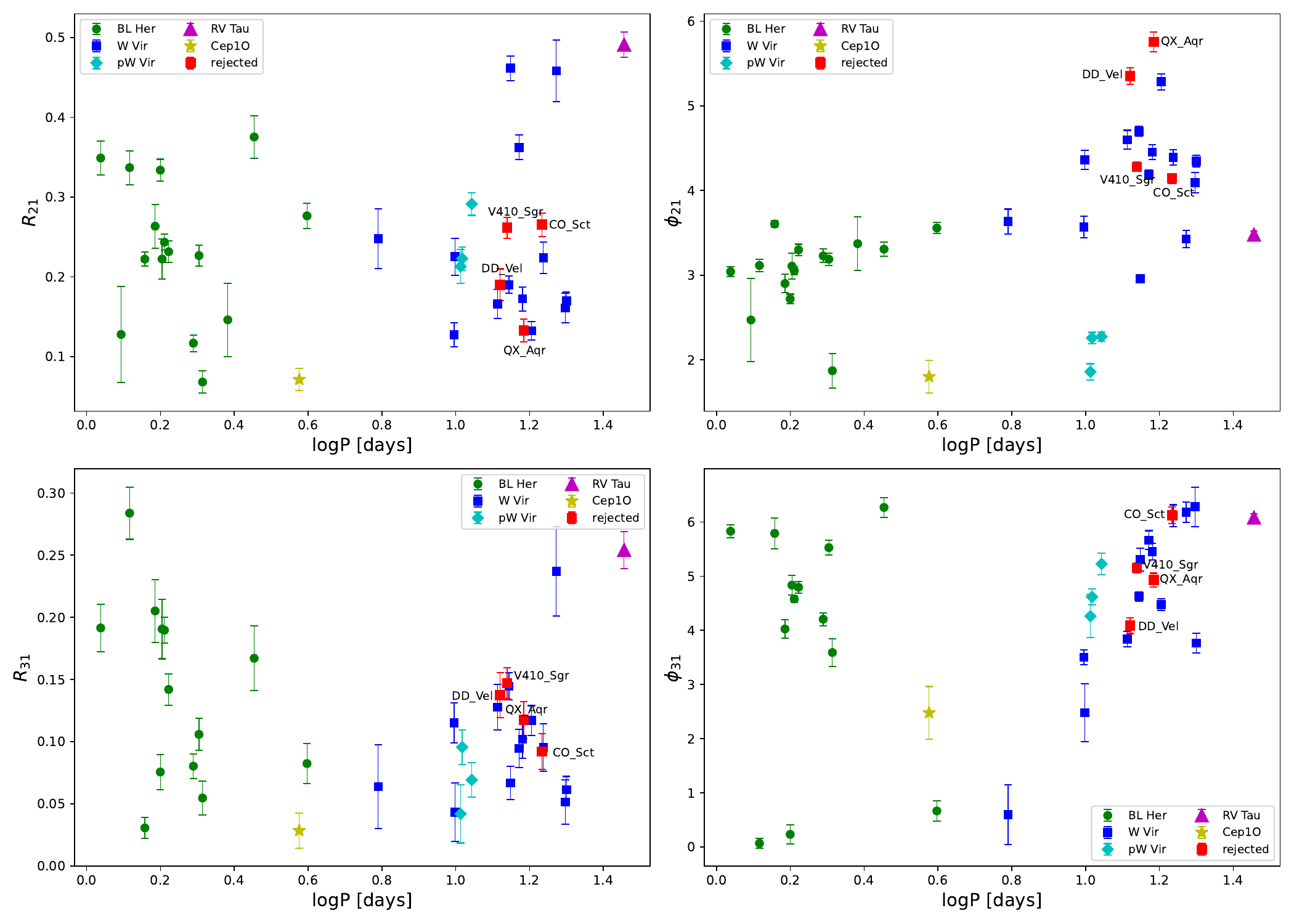}
   \caption{Fourier parameters from the analysis of $g_{P1}$-band light curves of stars from 
            our sample in function of literature periods. 
            }
      \label{fig:fp}
    \end{figure*}
%

\FloatBarrier 
\twocolumn

%

\begin{sidewaystable*}
\section{Sample of Type~II Cepheids.} 
\label{app:sum}
Table~\ref{tab:t2cep} summarizes the parameters of all MW Type~II Cepheids from our sample. 

\scriptsize
\caption{\label{tab:t2cep} Sample of the MW Type~II Cepheids and their main parameters.}
\begin{tabular}{l@{\extracolsep{2.5em}}llllcccccccc}
\hline\hline 
Star & Type & Period & $\varpi_{DR3}$ & RUWE & GOF & E(B-V) & 
$<g>$ & N$g$ & $<r>$ & N$r$ & $<i>$ & N$i$ \\
     &  & (days) & (mas)  &  &  & (mag) & (mag) & & (mag) & & (mag) & \\ 
\hline
\multicolumn{13}{l}{BL~Her type stars} \\ 
\hline
AU~Peg    & BLHer & 2.4121 & 1.6771 $\pm$ 0.0200 & 1.25 & 6.79 & 0.0460 $\pm$ 0.0013 & 9.585 $\pm$ 0.005 & 52 & 9.041 $\pm$ 0.003 & 51 & 8.868 $\pm$ 0.003 & 50 \\
BL~Her    & BLHer & 1.307445 & 0.8705 $\pm$ 0.0179 & 1.29 & 10.20 & 0.0670 $\pm$ 0.0018 & 10.326 $\pm$ 0.006 & 46 & 10.110 $\pm$ 0.004 & 45 & 10.058 $\pm$ 0.003 & 46 \\
BX~Del    & BLHer & 1.091787 & 0.3860 $\pm$ 0.0150 & 1.13 & 3.20 & 0.1000 $\pm$ 0.0059 & 12.386 $\pm$ 0.004 & 53 & 12.097 $\pm$ 0.004 & 52 & 12.014 $\pm$ 0.005 & 50 \\
FM~Del    & BLHer & 3.954 & 0.2498 $\pm$ 0.0135 & 0.93 & -1.71 & 0.0870 $\pm$ 0.0013 & 12.636 $\pm$ 0.004 & 50 & 12.253 $\pm$ 0.003 & 51 & 12.079 $\pm$ 0.003 & 48 \\
NW~Lyr    & BLHer & 1.6011823 & 0.3444 $\pm$ 0.0122 & 1.16 & 4.20 & 0.1200 $\pm$ 0.0034 & 12.649 $\pm$ 0.010 & 23 & 12.305 $\pm$ 0.006 & 23 & 12.188 $\pm$ 0.006 & 23 \\ 
RT~TrA     & BLHer & 1.9461124 & 1.0404 $\pm$ 0.0162 & 0.95 & -1.83 & 0.1120 $\pm$ 0.0046 & 10.033 $\pm$ 0.003 & 48 & 9.626 $\pm$ 0.002 & 48 & 9.518 $\pm$ 0.002 & 47 \\ 
SW~Tau     & BLHer & 1.583535 & 1.2554 $\pm$ 0.0222 & 1.25 & 5.52 & 0.2520 $\pm$ 0.0093 & 9.996 $\pm$ 0.004 & 50 & 9.584 $\pm$ 0.003 & 47 & 9.428 $\pm$ 0.003 & 50 \\ 
V439~Oph  & BLHer & 1.89303 & 0.5105 $\pm$ 0.0163 & 1.17 & 4.68 & 0.2680 $\pm$ 0.0049 & 12.509 $\pm$ 0.019 & 15 & 11.908 $\pm$ 0.011 & 15 & 11.614 $\pm$ 0.010 & 15 \\
V465~Oph$^{a}$  & BLHer & 2.8438 & 0.1776 $\pm$ 0.0254 & 1.44 & 9.71 & 0.3970 $\pm$ 0.0096 & 13.772 $\pm$ 0.008 & 46 & 13.122 $\pm$ 0.005 & 46 & 12.808 $\pm$ 0.006 & 46 \\ 
V477~Oph  & BLHer & 2.01572 & 0.2059 $\pm$ 0.0154 & 1.10 & 2.36 & 0.1680 $\pm$ 0.0023 & 14.075 $\pm$ 0.004 & 42 & 13.667 $\pm$ 0.004 & 42 & 13.473 $\pm$ 0.004 & 41 \\ 
V553~Cen  & BLHer & 2.06055 & 1.7605 $\pm$ 0.0224 & 0.94 & -0.97 & 0.0590 $\pm$ 0.0019 & 8.714 $\pm$ 0.003 & 172 & 8.315 $\pm$ 0.003 & 160 & 8.204 $\pm$ 0.003 & 158 \\ 
V606~Pup  & BLHer & 1.43775 & 0.5769 $\pm$ 0.0123 & 0.93 & -1.91 & 0.1690 $\pm$ 0.0157 & 11.518 $\pm$ 0.002 & 63 & 11.158 $\pm$ 0.002 & 64 & 11.076 $\pm$ 0.002 & 64 \\ 
V971~Aql  & BLHer & 1.62454 & 0.4836 $\pm$ 0.0219 & 1.26 & 5.10 & 0.1750 $\pm$ 0.0045 & 12.229 $\pm$ 0.003 & 60 & 11.785 $\pm$ 0.002 & 59 & 11.597 $\pm$ 0.002 & 59 \\ 
V2022~Sgr & BLHer & 1.533171 & 0.3562 $\pm$ 0.0200 & 1.18 & 3.72 & 0.3250 $\pm$ 0.0075 & 13.832 $\pm$ 0.006 & 37 & 13.285 $\pm$ 0.005 & 36 & 13.035 $\pm$ 0.004 & 38 \\
VY~Pyx    & BLHer & 1.23995 & 3.9710 $\pm$ 0.0186 & 0.89 & -3.57 & 0.0480 $\pm$ 0.0034 & 7.514 $\pm$ 0.007 & 69 & 7.102 $\pm$ 0.006 & 60 & 7.157 $\pm$ 0.004 & 62 \\
VZ~Aql    & BLHer & 1.668261 & 0.2797 $\pm$ 0.0162 & 1.06 & 1.13 & 0.3170 $\pm$ 0.0067 & 13.909 $\pm$ 0.003 & 55 & 13.347 $\pm$ 0.003 & 55 & 13.120 $\pm$ 0.002 & 55 \\
\hline
\multicolumn{13}{l}{W~Vir type stars} \\ 
\hline
AL~Lyr$^{a}$ & WVir & 12.978 & 0.2593 $\pm$ 0.0225 & 1.86 & 25.00 & 0.2940 $\pm$ 0.0116 & 12.500 $\pm$ 0.005 & 54 & 11.634 $\pm$ 0.004 & 52 & 11.244 $\pm$ 0.003 & 54 \\
AX~Tel     & WVir & 9.896 & 0.1511 $\pm$ 0.0160 & 1.15 & 3.53 & 0.0720 $\pm$ 0.0010 & 12.930 $\pm$ 0.003 & 46 & 12.425 $\pm$ 0.003 & 48 & 12.205 $\pm$ 0.003 & 47 \\
CO~Pup     & WVir & 16.042 & 0.3677 $\pm$ 0.0166 & 1.31 & 9.29 & 0.1570 $\pm$ 0.0032 & 11.205 $\pm$ 0.004 & 61 & 10.558 $\pm$ 0.004 & 61 & 10.270 $\pm$ 0.003 & 60 \\
CO~Sct     & WVir & 17.13799$^{*}$ & 0.1115 $\pm$ 0.0300 & 0.97 & -0.59 & 0.6590 $\pm$ 0.0209 & 14.812 $\pm$ 0.011 & 47 & 13.580 $\pm$ 0.009 & 47 & 12.937 $\pm$ 0.006 & 47 \\
DD~Vel     & WVir & 13.198 & 0.4742 $\pm$ 0.0127 & 1.08 & 2.13 & 1.3340 $\pm$ 0.0870 & 13.176 $\pm$ 0.004 & 50 & 11.902 $\pm$ 0.004 & 50 & 11.201 $\pm$ 0.003 & 50 \\
FI~Sct     & WVir & 14.85124 & 0.1594 $\pm$ 0.0209 & 1.07 & 1.31 & 0.8790 $\pm$ 0.0740 & 14.916 $\pm$ 0.007 & 63 & 13.646 $\pm$ 0.004 & 63 & 12.968 $\pm$ 0.004 & 63 \\
HQ~Car     & WVir & 14.067 & 0.2132 $\pm$ 0.0105 & 0.98 & -0.56 & 0.4190 $\pm$ 0.0317 & 12.701 $\pm$ 0.004 & 46 & 12.076 $\pm$ 0.003 & 46 & 11.782 $\pm$ 0.004 & 46 \\
HQ~Cen     & WVir & 9.9585$^{**}$ & 0.3951 $\pm$ 0.0245 & 1.33 & 10.60 & 0.3350 $\pm$ 0.0130 & 12.289 $\pm$ 0.004 & 39 & 11.429 $\pm$ 0.003 & 39 & 11.042 $\pm$ 0.003 & 39 \\
MR~Ara     & WVir & 19.814$^{*}$ & 0.2356 $\pm$ 0.0236 & 0.97 & -0.75 & 0.1100 $\pm$ 0.0025 & 11.694 $\pm$ 0.009 & 60 & 11.174 $\pm$ 0.006 & 60 & 10.934 $\pm$ 0.005 & 60 \\
QX~Aqr     & WVir & 15.285 & 0.3097 $\pm$ 0.0245 & 1.31 & 5.89 & 0.1030 $\pm$ 0.0061 & 12.547 $\pm$ 0.004 & 55 & 11.991 $\pm$ 0.003 & 55 & 11.738 $\pm$ 0.004 & 55 \\
RS~Pav     & WVir & 19.945$^{**}$ & 0.3797 $\pm$ 0.0157 & 0.99 & 5.89 & 0.0720 $\pm$ 0.0012 & 10.929 $\pm$ 0.005 & 59 & 10.281 $\pm$ 0.004 & 59 & 10.011 $\pm$ 0.004 & 59 \\
ST~Pup$^{a}$ & WVir & 18.73 & 0.4233 $\pm$ 0.0232 & 2.07 & 25.61 & 0.1200 $\pm$ 0.0019 & 10.206 $\pm$ 0.017 & 37 & 9.829 $\pm$ 0.011 & 37 & 9.623 $\pm$ 0.008 & 37 \\
TX~Del$^{a}$ & WVir & 6.165907 & 0.9413 $\pm$ 0.0294 & 1.94 & 27.90 & 0.0850 $\pm$ 0.0014 & 9.492 $\pm$ 0.008 & 48 & 9.000 $\pm$ 0.007 & 47 & 8.863 $\pm$ 0.005 & 46 \\
V410~Sgr     & WVir & 13.7753 & 0.0746 $\pm$ 0.0186 & 1.39 & 7.93 & 0.1940 $\pm$ 0.0047 & 13.184 $\pm$ 0.005 & 44 & 12.508 $\pm$ 0.004 & 44 & 12.205 $\pm$ 0.004 & 44 \\
V741~Sgr     & WVir & 15.16817 & 0.2372 $\pm$ 0.0208 & 0.93 & -1.35 & 0.2850 $\pm$ 0.0082 & 13.144 $\pm$ 0.006 & 49 & 12.425 $\pm$ 0.004 & 49 & 12.100 $\pm$ 0.003 & 49 \\
VZ~Tau     & WVir & 13.943 & 0.2817 $\pm$ 0.0225 & 1.38 & 8.16 & 0.4830 $\pm$ 0.0354 & 13.090 $\pm$ 0.004 & 47 & 12.167 $\pm$ 0.003 & 46 & 11.699 $\pm$ 0.003 & 47 \\
W~Vir     & WVir & 17.2736 & 0.5082 $\pm$ 0.0222 & 1.06 & 2.13 & 0.0360 $\pm$ 0.0009 & 10.209 $\pm$ 0.009 & 53 & 9.756 $\pm$ 0.005 & 53 & 9.562 $\pm$ 0.006 & 53 \\ 
\hline
\multicolumn{13}{l}{other variable types} \\ 
\hline
V572~Aql  & Cep1O & 3.76733 & 0.1391 $\pm$ 0.0178 & 1.13 & 3.68 & 0.1510 $\pm$ 0.0030 & 11.454 $\pm$ 0.002 & 50 & 10.956 $\pm$ 0.002 & 49 & 10.769 $\pm$ 0.002 & 48 \\
AL~Vir     & pWVir & 10.3065 & 0.4909 $\pm$ 0.0190 & 0.98 & -0.32 & 0.0720 $\pm$ 0.0007 & 9.745 $\pm$ 0.005 & 26 & 9.397 $\pm$ 0.005 & 26 & 9.261 $\pm$ 0.005 & 26 \\
AP~Her     & pWVir & 10.411 & 0.3866 $\pm$ 0.0152 & 1.13 & 3.74 & 0.3740 $\pm$ 0.0191 & 11.127 $\pm$ 0.005 & 54 & 10.575 $\pm$ 0.005 & 54 & 10.304 $\pm$ 0.004 & 55 \\
BH~Oph     & pWVir & 11.048 & 0.1714 $\pm$ 0.0172 & 1.04 & 0.82 & 0.1390 $\pm$ 0.0089 & 12.281 $\pm$ 0.006 & 59 & 11.871 $\pm$ 0.004 & 58 & 11.694 $\pm$ 0.003 & 57 \\
TW~Cap     & RVTau & 28.585 & 0.2000 $\pm$ 0.0193 & 1.26 & 6.17 & 0.0850 $\pm$ 0.0024 & 10.774 $\pm$ 0.014 & 61 & 10.379 $\pm$ 0.008 & 61 & 10.181 $\pm$ 0.006 & 61 \\
V1711~Sgr$^{a}$ & RVTau & 28.46$^{*}$ & 0.2218 $\pm$ 0.0254 & 1.52 & 11.60 & 0.1060 $\pm$ 0.0038 & 10.871 $\pm$ 0.015 & 44 & 10.448 $\pm$ 0.010 & 44 & 10.263 $\pm$ 0.008 & 44 \\
\hline
\multicolumn{13}{@{}l@{}}{\parbox{0.96\textwidth}{
   star: name of a~T2Cep; 
   type: type of variability after reclassification (BLHer: BL~Her type star, WVir: W~Vir type 
   star, pWVir: peculiar W~Vir type star, Cep1O: first overtone mode classical Cepheid, RVTau: 
   RV~Tau type star); 
   period: period of a~given star adopted from AAVSO database; 
   $\varpi_{DR3}$: parallax from Gaia~DR3 catalog corrected with \citet{LindegrenBastian2021} 
   corrections; 
   RUWE: renormalized unit weight error from the Gaia~DR3 catalog; 
   GOF: goodness-of-fit from Gaia~DR3 catalog; 
   E(B-V): reddening value from \citet{SF2011} reddening map corrected for the MW model 
   by \citet{DrimmelSpergel2001}; 
   $<g>$, $<r>$, $<i>$: mean magnitude from Fourier series fitting for the Pan-STARRS 
   $g_{P1}r_{P1}i_{P1}$ filters, respectively. \\ 
   N$g$, N$r$, N$i$: number of points. \\
   $^{a}$ Stars rejected based on the RUWE and GOF parallax quality parameters given in the 
   Gaia~DR3 catalog. \\ 
   $^{*}$ Mean magnitudes calculated for doubled period. \\ 
   $^{**}$ Half of the period from AAVSO, but the mean magnitudes were calculated for double the 
   period value given here.. 
}}
\end{tabular}
\end{sidewaystable*}

\FloatBarrier 
\twocolumn

\onecolumn
\section{The Sloan band light curves of Galactic Type~II Cepheid stars analazyed in this work.} 
\label{app:lc}

Figure~\ref{fig:fig1} presents the Sloan $g_{P1}r_{P1}i_{P1}$ bands light curves of $16$ BL~Her 
type stars used in this study, while Fig.~\ref{fig:fig2} shows same light curves for 
$17$ W~Vir 
type stars. Figure~\ref{fig:fig3} shows the light curves of other types of variables classified 
in this work (one first-overtone Type~I Cepheid, three pW~Vir and two RV~Tau). 
The presented light curves are available at the webpage of the Araucaria
Project: https://araucaria.camk.edu.pl/ and the CDS.

\begin{figure*}[h]
    \centering
    \includegraphics[width=\textwidth]{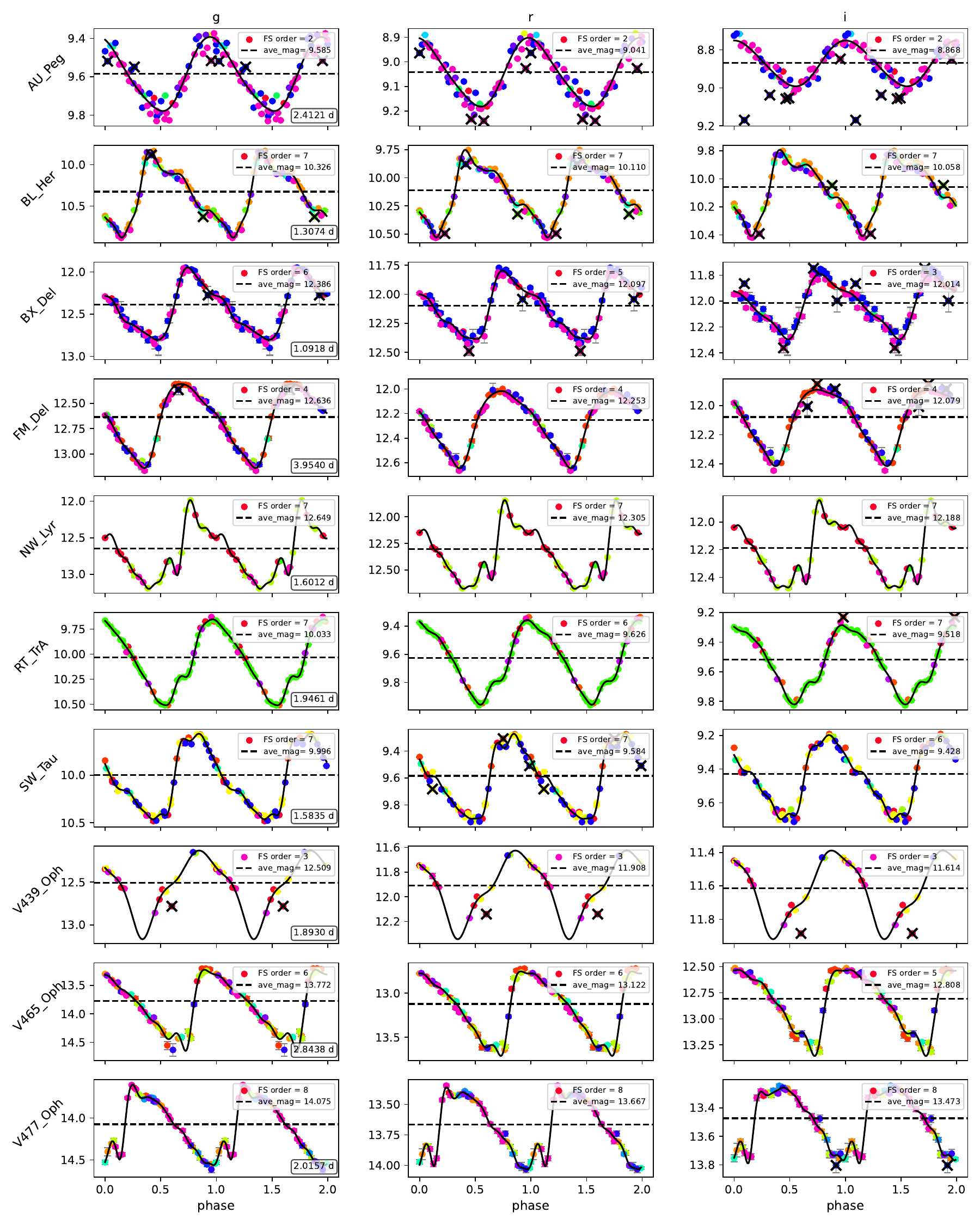}
    \caption{\label{fig:fig1} 
    Sloan--Pan-STARRS $g_{P1}r_{P1}i_{P1}$ band light curves of T2Cep stars of BL~Her 
    type analyzed in this work. 
    Horizontal dashed, black lines correspond to the determined mean magnitudes. 
    Different colors of points mark different telescopes used during the data collection, 
    while black crosses mark points rejected from the fitting. 
    Black lines show the best fit Fourier series.}
\end{figure*}

\begin{figure*}
    \ContinuedFloat
    \centering
    \includegraphics[width=\textwidth]{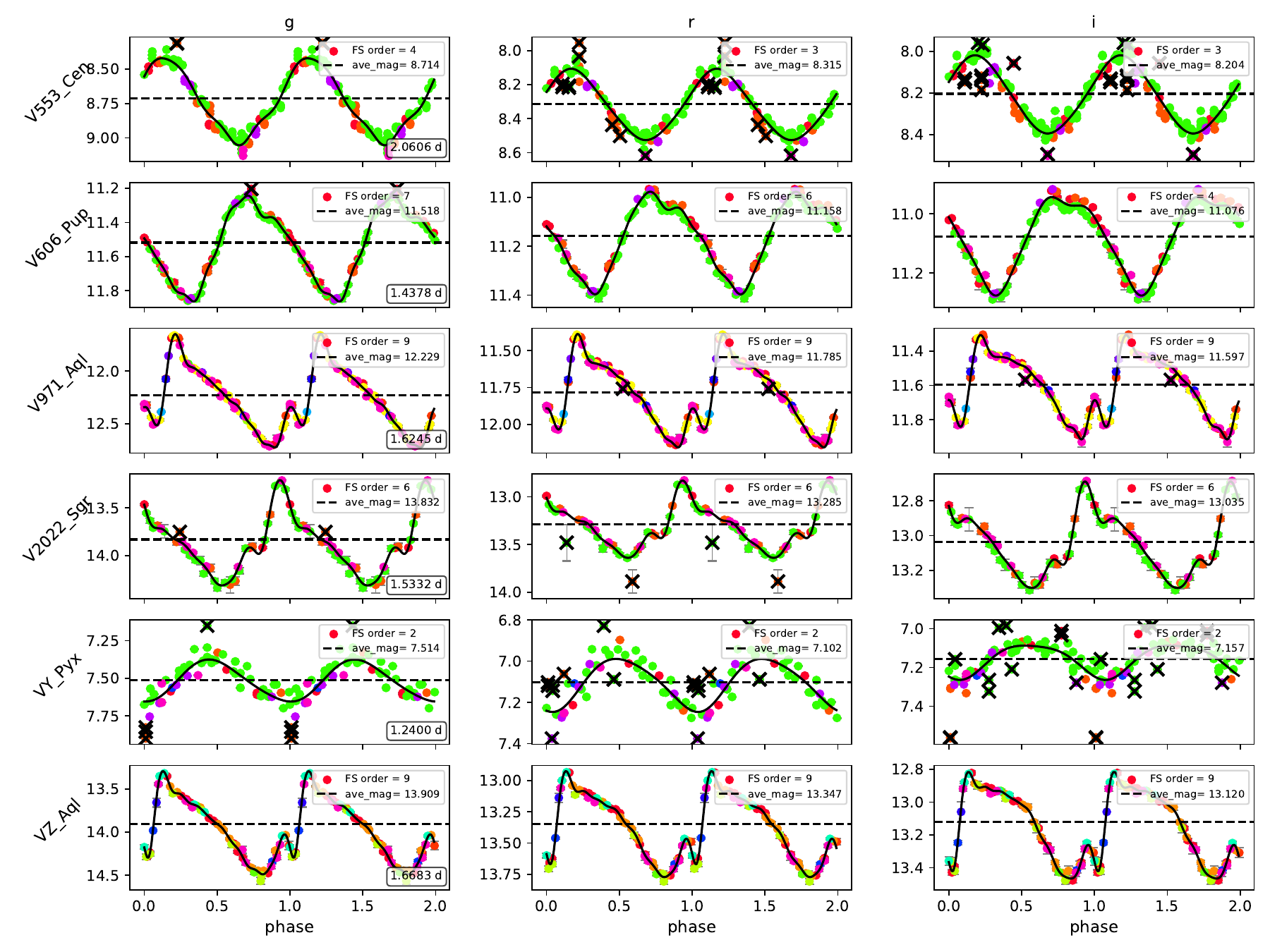}
    \caption{Continued from the previous page.}
\end{figure*}

\begin{figure*}[h]
    \centering
    \includegraphics[width=\textwidth]{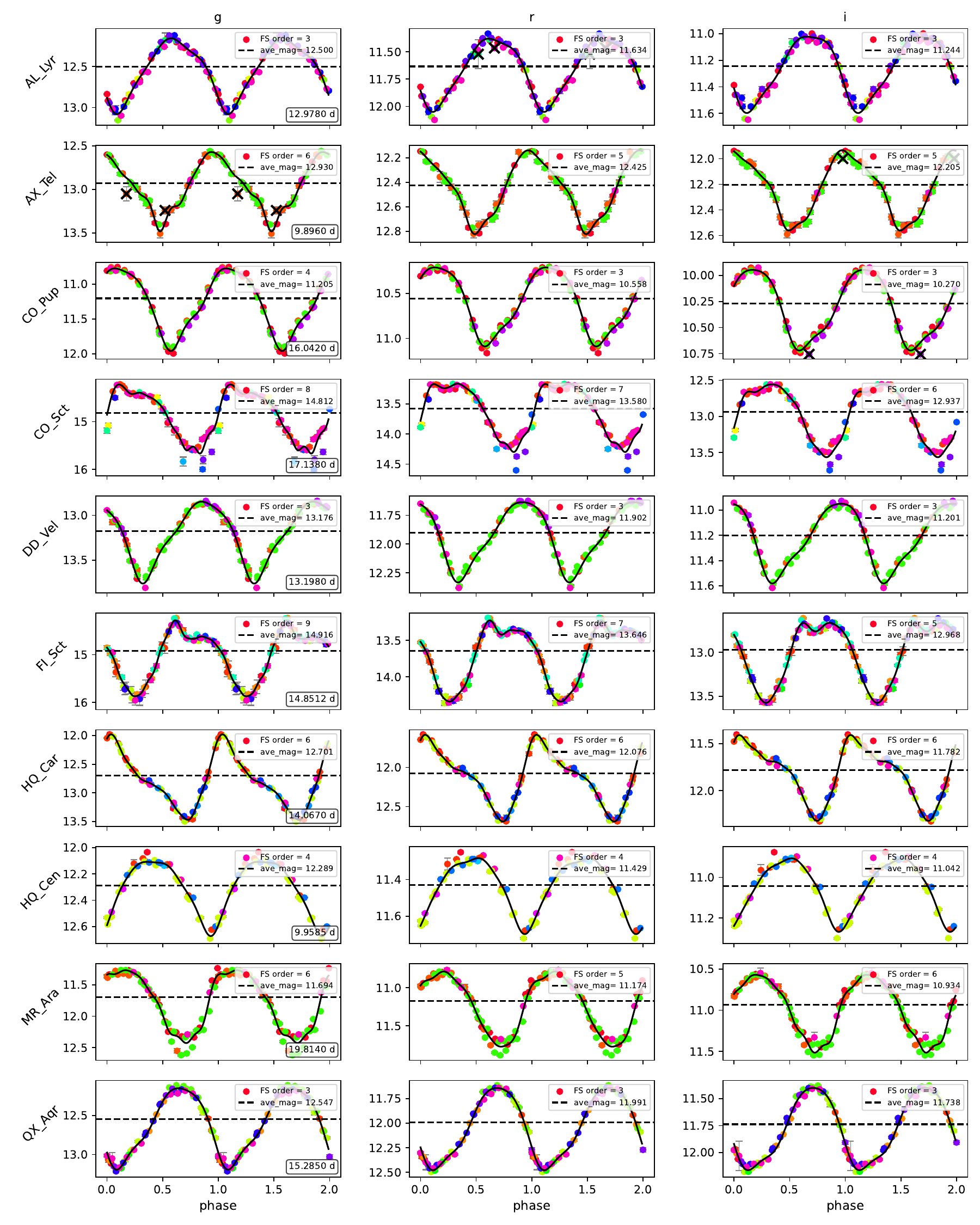}
    \caption{\label{fig:fig2} 
    Sloan--Pan-STARRS $g_{P1}r_{P1}i_{P1}$ band light curves of T2Cep stars of W~Vir type 
    analyzed in this work. 
    Horizontal dashed, black lines correspond to the determined mean magnitudes. 
    Different colors of points mark different telescopes used during the data collection, 
    while black crosses mark points rejected from the fitting. 
    Black lines show the best fit Fourier series.}
\end{figure*}

\begin{figure*}
    \ContinuedFloat
    \centering
    \includegraphics[width=\textwidth]{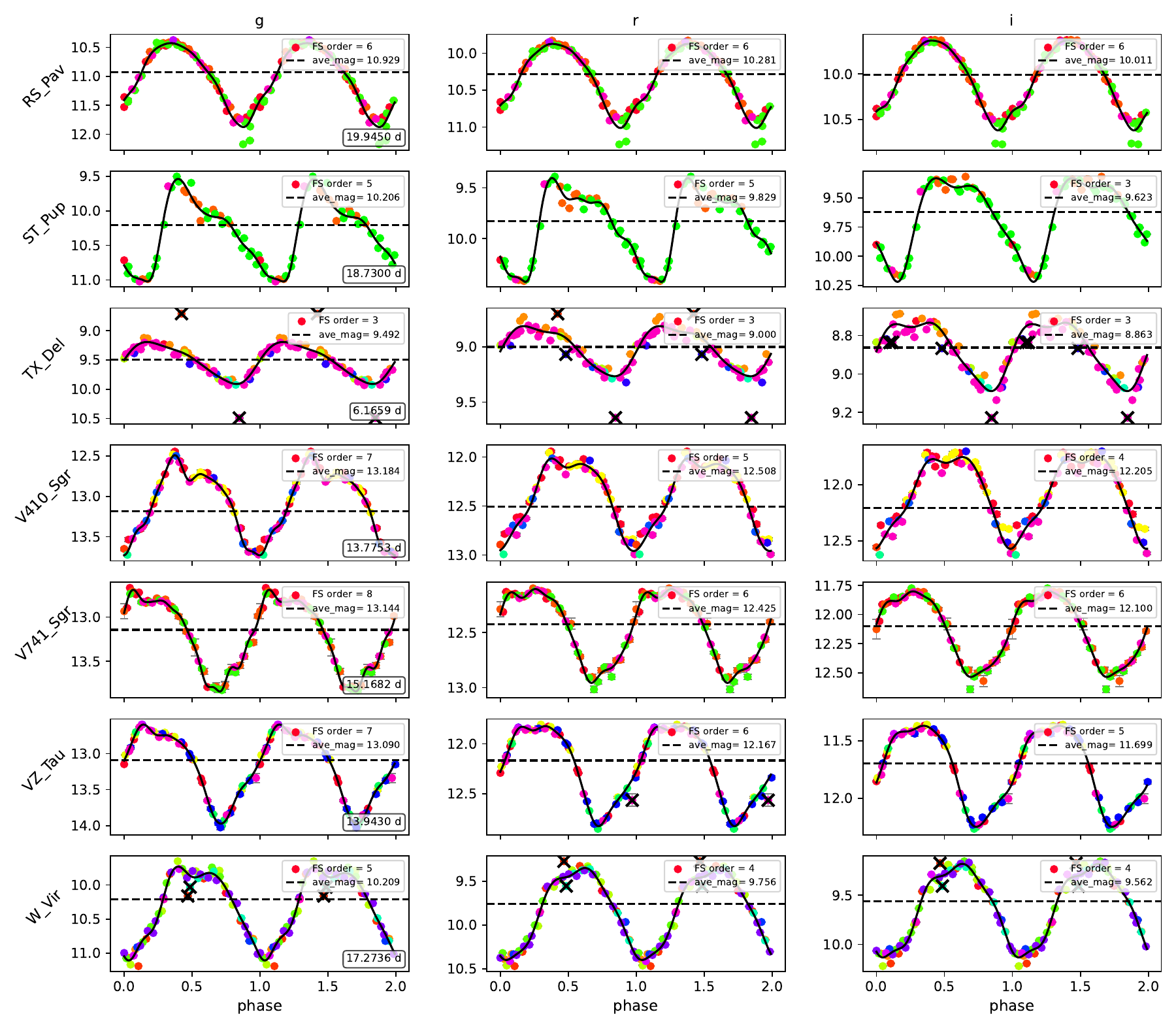}
    \caption{Continued from the previous page.}
\end{figure*}

\begin{figure*}[h]
    \centering
    \includegraphics[width=\textwidth]{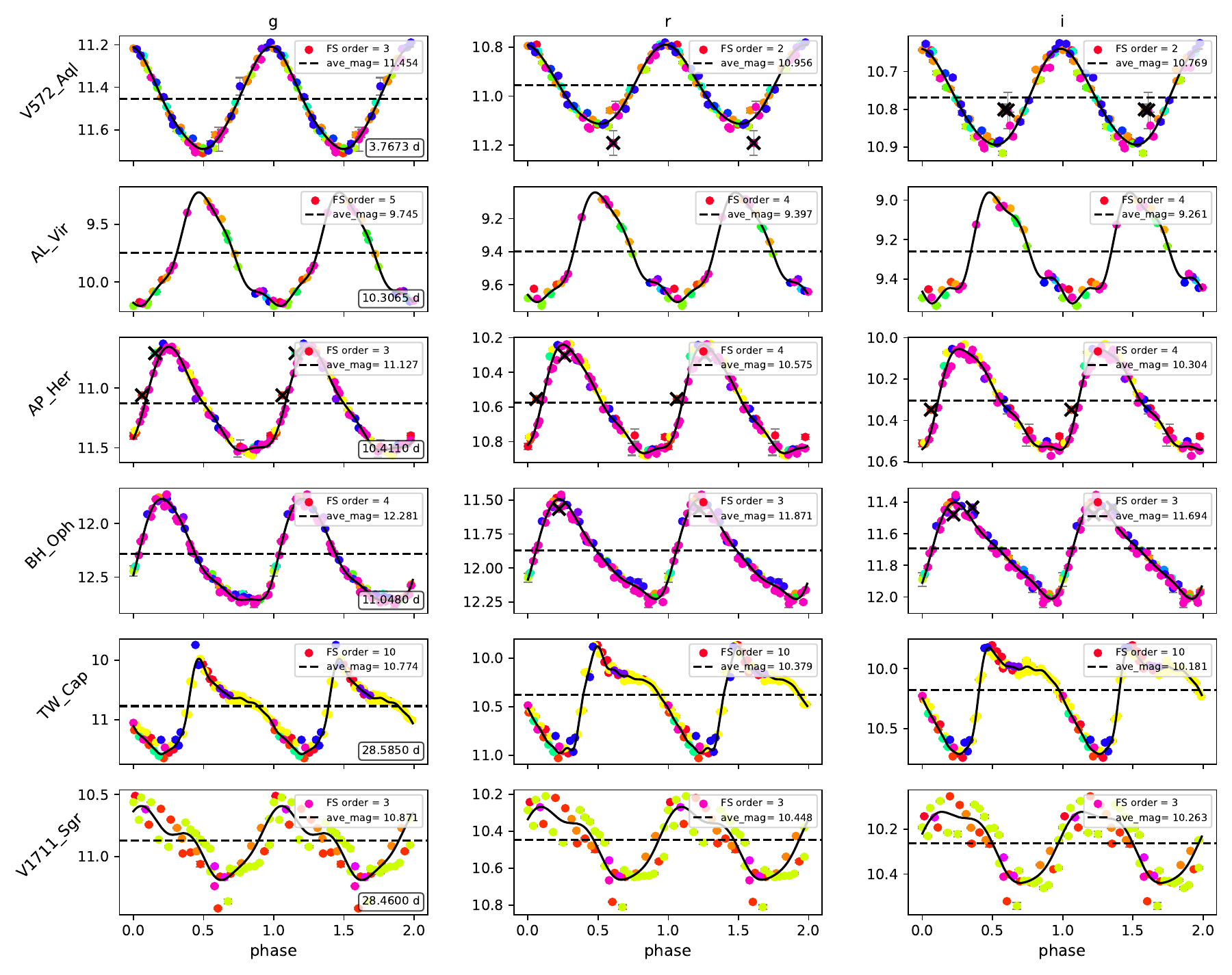}
    \caption{\label{fig:fig3} 
    Sloan--Pan-STARRS $g_{P1}r_{P1}i_{P1}$ band light curves of variables of other types 
    (Cepheid 1O, pW~Vir and RV~Tau) analyzed in this work. 
    Horizontal dashed, black lines correspond to the determined mean magnitudes. 
    Different colors of points mark different telescopes used during the data collection, 
    while black crosses mark points rejected from the fitting. 
    Black lines show the best fit Fourier series.}
\end{figure*}

\end{appendix}

\end{document}